\documentclass[%
 reprint,
 amsmath,amssymb,
 aps,
]{revtex4-2}

\usepackage{graphicx}
\usepackage{dcolumn}
\usepackage{bm}
\usepackage{geometry}
\usepackage{adjustbox}   

\usepackage[caption=false]{subfig}

\usepackage{longtable}   

\usepackage{float}
\usepackage{booktabs} 
\usepackage{hyperref}

\newgeometry{left=1.5cm,right=1.5cm,top=1.5cm,bottom=2cm}
\usepackage{booktabs} 

\begin{document}

\preprint{APS/123-QED}

\title{Learning Magnetic Order Classification from Large-Scale Materials Databases}

\author{Ahmed E. Fahmy}
 \email{abdelazim.2@osu.edu}
\affiliation{%
 Department of Physics, The Ohio State University, Columbus, OH 43210, USA
}%

\date{\today}

\begin{abstract}
The reliable identification of magnetic ground states remains a major challenge in high-throughput materials databases, where density functional theory (DFT) workflows often converge to ferromagnetic (FM) solutions. Here, we partially address this challenge by developing machine learning classifiers trained on experimentally validated MAGNDATA magnetic materials leveraging a limited number of simple compositional, structural, and electronic descriptors sourced from the Materials Project database. Our propagation vector classifiers achieve accuracies above $92\%$, outperforming a recent equivariant-neural-network study on a differently constructed dataset in reliably distinguishing zero from nonzero propagation vector structures, and exposing a systematic ferromagnetic bias inherent to the Materials Project database for more than $6,840$ candidate materials. In parallel, LightGBM and XGBoost models trained directly on the Materials Project labels achieve accuracies of $82\%$ and $85\%$, respectively (with macro $F_1$ average scores of $66\%$ and $63\%$), which proves useful for large-scale screening for magnetic classes, if refined by MAGNDATA-trained classifiers. These results underscore the role of machine learning techniques as corrective and exploratory tools, enabling more trustworthy databases and accelerating progress toward the identification of materials with various properties.
\end{abstract}

\maketitle


\section{Introduction}
Identifying the true magnetic ground state of a material remains a central challenge in condensed matter physics and materials discovery. On the computational side, ab initio methods such as density functional theory (DFT) have become indispensable for exploring magnetism, yet they face several well-documented limitations. The predicted magnetic state depends sensitively on the chosen exchange–correlation functional and on the inclusion of on-site Coulomb corrections through the Hubbard
U, which can drastically alter the relative stability of competing spin configurations \cite{Cohen_Science_2008,Himmetoglu_IJQC_2014}. Furthermore, the outcome of a DFT calculation is strongly influenced by the initialization of local magnetic moments. High-throughput workflows, such as those employed in large materials databases, typically adopt a ferromagnetic starting configuration, which systematically biases the results toward FM solutions and may overlook lower-energy antiferromagnetic, ferrimagnetic, or incommensurate ground states \cite{Jain_HandbookMatModeling_2018}. More sophisticated ab initio techniques—including DFT+DMFT (dynamical mean-field theory) \cite{Kotliar_RMP_2006}, hybrid functionals, and quantum Monte Carlo \cite{Foulkes_RMP_2001}—offer improved accuracy for correlated electron systems, but their high computational cost renders them impractical for systematic screening of large numbers of materials. Even with conventional DFT, reliably determining a material’s ground-state magnetic structure often requires exploring many candidate spin arrangements, sometimes guided by symmetry analysis or model Hamiltonians, and comparing their relative total energies \cite{Pickett_PRB_1989,Andersen_PRB_1977}. This procedure becomes prohibitively expensive for complex or low-symmetry systems, underscoring the difficulty of ab initio magnetic ground-state determination at scale.

Experimentally, techniques such as neutron scattering \cite{Lovesey_1986,Zaliznyak_Tranquada_2012} and, more recently, resonant inelastic X-ray scattering (RIXS) \cite{Ament_RMP_2011} and muon spin rotation (µSR) \cite{Blundell_CP_1999} have provided atomic-scale insights into static and dynamic magnetic structures. Despite their unparalleled resolution, these probes depend on access to large-scale facilities $-$ neutron reactors, spallation sources, synchrotron light sources, or muon sources $-$ and are constrained by limited beam time availability, sample size requirements, and facility capacity. Consequently, the pace of experimental magnetic structure determination has been modest: even the most comprehensive crystallographic repositories, such as MAGNDATA, contain only about $2,300$ fully resolved magnetic structures spanning both commensurate and incommensurate orders since the 1950s \cite{Gallego_2016a,Gallego_2016b}. While continuous advances in experimental techniques (e.g., higher-flux neutron sources \cite{Sears_Neutron_1992} and next-generation synchrotrons \cite{Eriksson_JSR_2014}) promise incremental improvements, a purely experimental mapping of magnetic materials is unlikely to keep pace with the rapidly growing demand for new magnetic compounds. This bottleneck strongly motivates the integration of complementary approaches, such as machine learning, to accelerate exploration.

Given the challenges in magnetic structure determination from both computational and experimental angles, recent research has turned toward machine learning as a powerful complementary strategy. For example, Ghosh et al. applied regression and classification models to high-quality DFT and experimental datasets of actinide compounds, achieving $\sim 76\%$ accuracy in ordering (paramagnetic, ferromagnetic, antiferromagnetic) using a random forest classifier \cite{Ghosh_PRMaterials_2020}. More recently, Merker et al. developed an equivariant Euclidean neural network that predicts both magnetic order and propagation vector directly from atomic coordinates $-$ reporting average accuracies of $\sim 77.8\%$ for magnetic order and $\sim 73.6\%$ for propagation vector classification, with an impressive $\sim 91\%$ accuracy for identifying non-magnetic structures \cite{Merker_iScience_2022}. These studies demonstrate that machine learning can effectively assist—or even partly replace—the conventional “guessing” step in DFT workflows, though fully circumventing first-principles calculations remains challenging.

A complete description of magnetism remains inherently complex \cite{RodriguezCarvajal_ComptesPhys_2019}. Here, we focus on two streamlined descriptors: magnetic order labels (FM, AFM, FiM, NM) and propagation vectors. Magnetic order labels reduce complexity into categories meaningful for applications (e.g., FM and FiM display net magnetization, AFM and NM do not), while propagation vectors capture more intricate ordering patterns when they deviate from zero. Although simplified, these descriptors offer a balanced compromise between expressiveness and tractability—serving our goal of scalable, interpretable predictions.

In this paper, our goal is to build ML classifiers that can predict the magnetic order (FM, FiM, AFM, NM) on the Materials Project (MP) database \cite{Jain_APLMaterials_2013,Jain_HandbookMatModeling_2018}, a binary information about the propagation vector (zero or nonzero) on Bilbao Crystallographic Server (BCS)'s MAGNDATA \cite{Gallego_2016a,Gallego_2016b}, and finally try to expose the FM bias on the MP database through the efficient MAGNDATA-trained classifier. Although these outputs do not cover a full description of the magnetic order, they still contain critical information on the existence of local magnetic moments, the magnitude of the net magnetization, and the relative spin alignments, all of which are highly informative for distinguishing between different magnetic classes.

\section{Data Collection \& Pre-processing}
\label{sec:data collecting}
 Our strategy is to use a few simple descriptors that are linked either to the formation of magnetic moments or the interactions between them, which eventually lead to the formation of the long-range magnetic order. Therefore, we consider three classes of descriptors: compositional, structural, and electronic. The compositional part contains the different constituent chemical elements that form the compound, and therefore implicitly encoding information about their likely oxidation states and electronic configurations, which in turn affect the likelihood of forming local magnetic moments. The chemical composition was implemented using a one-hot encoding scheme, where each element is represented by a unique basis vector in a high-dimensional binary space. The structural part consists of the crystal system (cubic, tetragonal, orthorhombic, hexagonal, trigonal, monoclinic, and triclinic), the atomic density, volume of the unit cell (which, combined with the atomic density, relates to the number density), and mass density, which is indirectly related to the atomic masses of the constituent elements. These features are linked to the symmetry of the crystal, the packing and compactness of the structure, and the lattice geometry, which in turn influence the superexchange and direct exchange pathways that govern magnetic interactions.

Electronic descriptors — such as the band gap, conduction band minimum (CBM), valence band maximum (VBM), and Fermi energy — supplement this structural information by characterizing the degree of electron localization and the availability of itinerant carriers. In particular, materials with large band gaps and partially filled d- or f-shells typically exhibit localized electrons, which favor the formation of robust local moments and superexchange-driven antiferromagnetic or ferrimagnetic order \cite{anderson1950,Goodenough1963,kanamori1959}. 
In contrast, metallic systems often support itinerant magnetism \cite{stoner1938,kubler2021} or spin-density-wave states \cite{overhauser1962,fawcett1988}.
Exceptions exist, including correlated systems where local and itinerant behavior coexist \cite{mott1990,coleman2007}. By linking electron localization and carrier availability to spin interactions, these descriptors are essential predictors of the type and stability of magnetic order in crystalline materials. It is worth noting that some of the descriptors we adopt do not give useful information unless combined with other features such as VBM, CBM, Fermi energy or volume. In keeping with the goal of demonstrating that even a coarse-grained descriptor set is sufficient to train our classifiers, we deliberately restrict our feature set to the descriptors discussed above. Other meaningful descriptors (e.g. electronegativity and atomic dipole polarizability that are considered in Ref. \cite{Merker_iScience_2022}) are intentionally excluded to keep the pipeline as simple as possible and strictly defined by what an end-user can easily pull from the MP database itself without further pre-processing. Such descriptors could be related indirectly to the compositional and electronic features that we employ. To make this quantitative, we will later relax this restriction and quantify the effect of including such descriptors on classifier performance.

For the MP classifier, we construct a dataset of the descriptors and the magnetic class for each material through the Materials Project API. This results in $154,803$ materials, with $104,795$ unique ones as there are materials with more than one MP profile. Due to continuous updates of the MP, the number may change upon a new access. We focus on materials with at least one magnetic element such as transition metal elements (Sc, Ti, V, Cr, Mn, Fe, Co, Ni, Cu, Y, Nb, Mo, Ru, Rh, Re, Os, Ir, and Pt), lanthanides (Ce, Pr, Nd, Sm, Eu, Gd, Tb, Dy, Ho, Er, Tm, and Yb) and actinides (Th, U, Np, and Pu). The magnetic order classification on the MP is determined from spin-polarized DFT calculations and post-processed with the CollinearMagneticStructureAnalyzer in the pymatgen library. A magnetic class is assigned based on the total magnetization of the unit cell and the alignment of different spins. If all spins align with nonzero net magnetization, the compound is labelled FM, if spins anti-align with nonzero net magnetization, it is labelled FiM, and if total magnetization vanishes but the local moments survive, it is AFM. If both the total magnetization and the local moments vanish, then it it NM. Based on this scheme, a magnetic class is assigned to each material on the MP. 

\begin{figure}[]
\centering
\subfloat[]{%
  \includegraphics[width=0.36\textwidth]{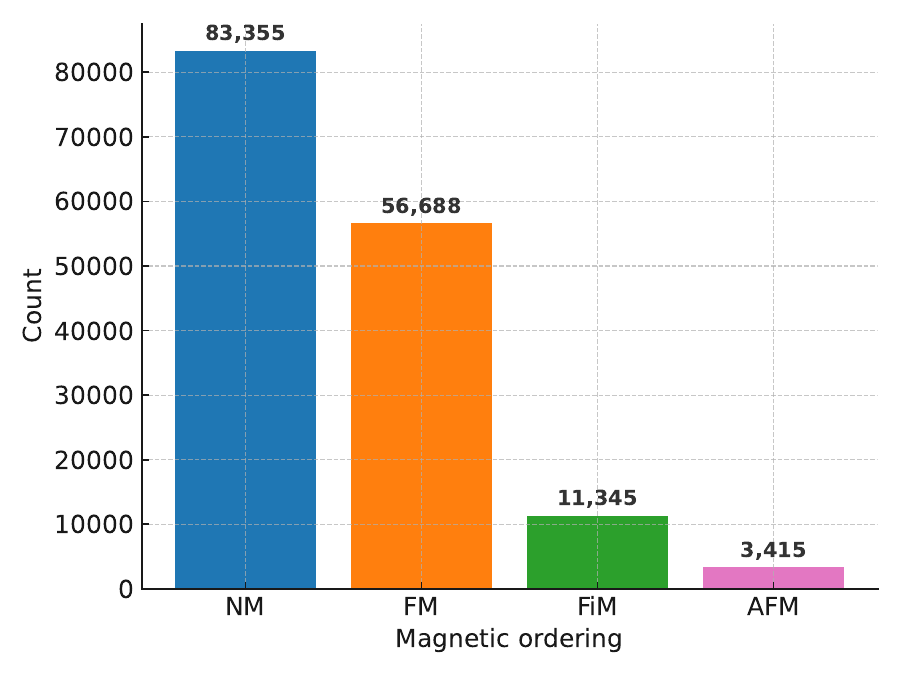}%
  \label{}}
\hfill
\subfloat[]{%
  \includegraphics[width=0.36\textwidth]{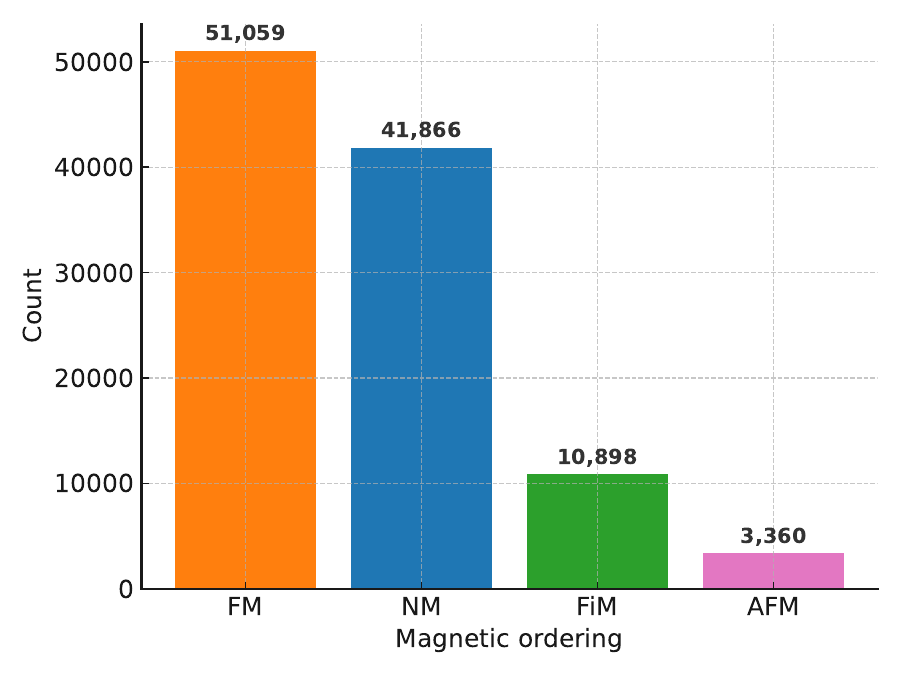}%
  \label{}}
\caption{(a) Distribution of magnetic classes in the Materials Project database without filtering. 
(b) Distribution of magnetic classes restricted to Materials Project compounds containing at least one magnetic element.}
\label{MaterialsProjectDistributionofclasses}
\end{figure}

\begin{figure}[]
\centering
\subfloat[]{%
  \includegraphics[width=0.47\linewidth]{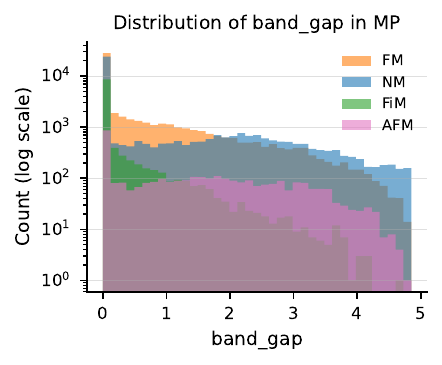}%
  \label{fig:dist-bandgap}}
\hfill
\subfloat[]{%
  \includegraphics[width=0.47\linewidth]{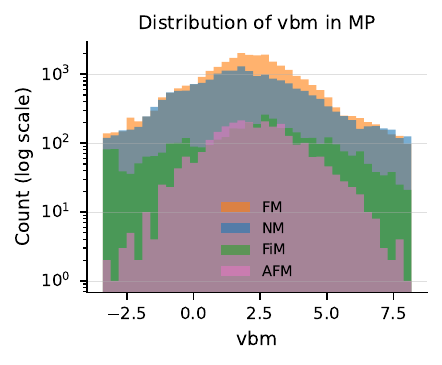}%
  \label{fig:dist-vbm}}
\\[1ex]
\subfloat[]{%
  \includegraphics[width=0.47\linewidth]{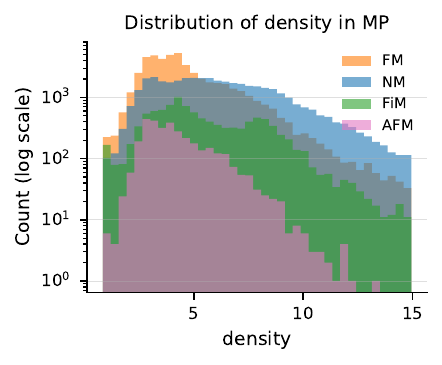}%
  \label{}}
\hfill
\subfloat[]{%
  \includegraphics[width=0.47\linewidth]{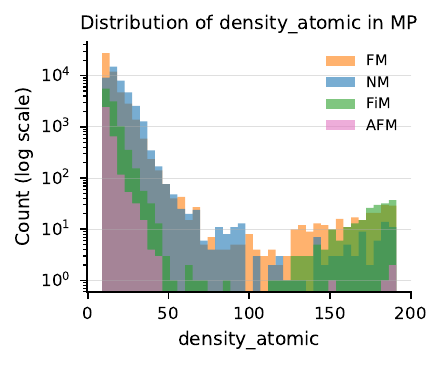}%
  \label{}}
\caption{Distributions (log-scale counts) of selected features in the refined Materials Project dataset, restricted to compounds containing at least one magnetic element and resolved by magnetic class: (a) electronic band gap, (b) valence band maximum (VBM), (c) mass density, and (d) atomic density. These descriptors highlight systematic differences between ferromagnetic (FM), antiferromagnetic (AFM), ferrimagnetic (FiM), and nonmagnetic (NM) compounds.}
\label{dataanalytics}
\end{figure}

The MP ordering is not derived from experiment but from the sign of the total magnetic moment in a single spin-polarized DFT calculation, initialized with a ferromagnetic guess (all MAGMOM values set nonzero and parallel). For compounds containing no magnetic ions, the self-consistent moment should relax to zero; in practice, however, a small residual moment often persists due to finite electronic smearing, incomplete k-point convergence, and the absence of an explicit antiferromagnetic reference calculation for comparison. When the magnitude of the total moment exceeds MP's internal threshold, the compound is classified as FM even when no physical mechanism for ferromagnetism exists. This is a documented limitation of MP's magnetic classification scheme — see Ref. \cite{Horton2019MP}, which describes the MP magnetic workflow in detail and reports that the concordance between MP magnetic ordering labels and experimental MAGNDATA classifications is limited, with compounds of weak or numerically-induced moments constituting a known class of mislabels. Finally, the magnetic class labels depend mainly on the used DFT functional (GGA or GGA+U) which can be sensitive the value of the Hubbard $U$. As a result of these limitations, the MP dataset shows high bias towards the FM and NM classes while weakly capturing FiM and AFM classes, as shown in the histogram distribution in Fig. \ref{MaterialsProjectDistributionofclasses}. Before adding the filter of the existence of magnetic elements, the NM class dominates with the FM class right behind it, while the filter produces an opposite distribution, as shown in Fig. \ref{MaterialsProjectDistributionofclasses} (b). The different magnetic classes of the filtered set of materials shows some distinct distributions over the selected descriptors as presented in Fig. \ref{dataanalytics} where we select two electronic features: the band gap and the valence band maximum, and two structural features: the density and the atomic density. Therefore, while useful for a large-scale screening, the magnetic class labels on the MP should be treated as heuristic rather than conclusive regarding describing the true magnetic ground state. \\

\begin{figure}[]
\centering
\subfloat[]{%
  \includegraphics[width=0.36\textwidth]{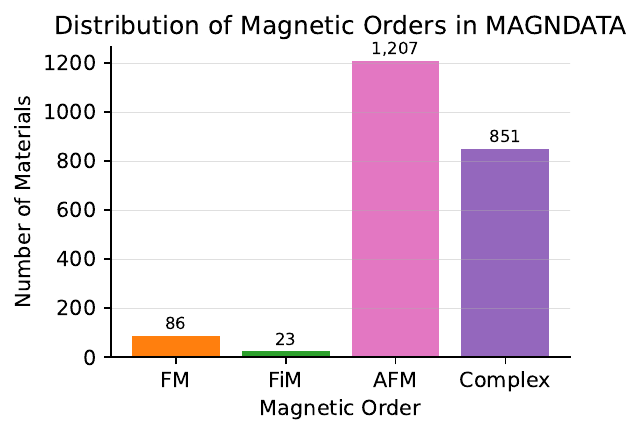}%
  \label{}}
\hfill
\subfloat[]{%
  \includegraphics[width=0.36\textwidth]{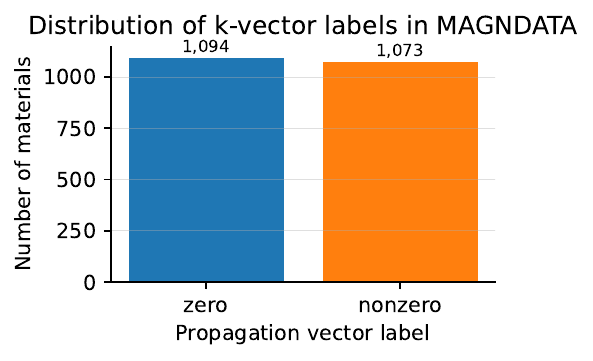}%
  \label{}}
\caption{Distributions of magnetic classes in the MAGNDATA dataset: 
(a) magnetic order labels, including ferromagnetic (FM), ferrimagnetic (FiM), antiferromagnetic (AFM), and complex orders; 
(b) propagation vector labels, distinguishing between zero and nonzero $k$-vectors. These distributions highlight the prevalence of AFM and complex orders in experimentally determined magnetic structures, as well as the near balance between commensurate and incommensurate propagation vectors.}
\label{MAGNDATAdistribution}
\end{figure}

Our second database, MAGNDATA on the BCS, is the most comprehensive experimentally-derived magnetic structures database through neutron diffraction studies. Although these experiments provide reliable description of magnetic orders, they are difficult to perform over a large scale and therefore it contains only $2,167$ commensurate magnetic materials (at the time of the data collection). On the contrary of the MP dataset, the distribution of the magnetic classes in MAGNDATA shows the dominance of the AFM with $1,207$ collinear AFM and $851$ non-collinear AFM, as shown in Fig. \ref{MAGNDATAdistribution} (a). This extreme imbalance makes it difficult to train a ML model given the rare data points about the FM and FiM classes. Therefore, since the distribution of the magnetic propagation vector (zero vs. nonzero) is balanced, as shown in Fig. \ref{MAGNDATAdistribution} (b), we choose it as a useful piece of information about the magnetic structure to train our ML models over. Note that we merged materials of more than one propagation vector into the nonzero class for simplicity to train a binary classifier over a balanced dataset. Adopting the same set of descriptors, we supplement the MAGNDATA dataset with the corresponding features from the MP database. This results in $3,980$ entries, with redundancy stemming from the existence of more than one MP-profile for a number of materials.

\paragraph*{Data flow and leakage controls.} In the rest of the paper, we use two distinct classifier families, each with its own scope. The MP classifiers are trained and evaluated on the MP database (more details in the next section), and are only applied to the MP database. Since training, evaluation, and application all come from the same database, there is no cross-database leakage to control for, and MAGNDATA entries are not removed from the MP pipeline. The MAGNDATA classifiers are trained and evaluated on MAGNDATA itself; the reported performance metrics measure generalization within MAGNDATA. We then apply these MAGNDATA-trained classifiers to MP to flag entries whose FM label are inconsistent with the learned propagation-vector predictions. This cross-database step is where leakage would matter: some MP entries are also in MAGNDATA, so applying the classifiers to them would amount to re-predicting materials the models had already seen during training. To prevent this, we remove all MAGNDATA materials from MP before the cross-database application, so the flagged list contains only MP entries that were never part of the MAGNDATA training data.

\section{Machine Learning Classifiers}
Now, we train ML classifiers for each of our datasets independently after partitioning each randomly with a stratified strategy (to take into account the imbalance in the MP dataset) into training, validation, and testing subsets with ratios of $70 \%$, $20 \%$, $10 \%$, respectively. Within the training set, we employed a multi-fold stratified cross validation (CV) to tune hyperparameters and to obtain a robust estimate of the model stability across the different folds. The separate validation set is used to compare the performance of the different trained models while the final test set is a final check set that is used only once for further confirmation of no overfitting (beside the model stability across the different folds in the CV employed during the training). It is worth noting that our ML classifier on the MAGNDATA propagation vector has two main goals. First is to get an accurate description of the propagation vector itself which reveals some information about the underlying magnetic structure. The second is to apply our most efficient propagation vector classifier to the MP dataset. This enables us to partially address the FM bias on the MP database through predicting a nonzero propagation vector, which excludes the possibility of a FM order. 

\paragraph*{Performance Metrics.} Throughout this work we evaluate the ML classifiers using three complementary metrics: accuracy, precision, and the $F_1$ score \cite{metrics1Powers2011,metrics2Sokolova2009}. For a given class, let TP (true positives) denote the number of materials correctly assigned to that class, FP (false positives) the number of materials incorrectly assigned to it, FN (false negatives) the number of materials that belong to the class but were assigned elsewhere, and TN (true negatives) the number of materials correctly assigned to another class. The first metric is accuracy which is the overall fraction of correct predictions,
\begin{equation}
\mathrm{Accuracy} = \frac{\mathrm{TP} + \mathrm{TN}}{\mathrm{TP} + \mathrm{TN} + \mathrm{FP} + \mathrm{FN}},
\label{eq:accuracy}
\end{equation}
and while intuitive, it is misleading on imbalanced datasets: a classifier that labels every material as FM would achieve high accuracy on an FM-heavy training set despite learning nothing useful. To avoid this, we rely more heavily on precision and recall, defined for each class as
\begin{equation}
\mathrm{Precision} = \frac{\mathrm{TP}}{\mathrm{TP} + \mathrm{FP}}, \qquad
\mathrm{Recall} = \frac{\mathrm{TP}}{\mathrm{TP} + \mathrm{FN}}.
\label{eq:precision_recall}
\end{equation}
Precision measures how often the classifier is correct when it flags a material as belonging to the class, while recall measures what fraction of the materials that actually belong to the class the classifier successfully identifies. The $F_1$ score combines these two through their harmonic mean,
\begin{equation}
F_1 = 2 \cdot \frac{\mathrm{Precision} \cdot \mathrm{Recall}}{\mathrm{Precision} + \mathrm{Recall}},
\label{eq:f1}
\end{equation}
and takes values between 0 (worst) and 1 (perfect classification). As a rough guide, a classifier that is not learning anything useful---for example, one that always predicts a single class---typically produces $F_1 \lesssim 1/k$ for a balanced $k$-class problem (e.g. around 0.25 for our four-class MP task and 0.5 for the binary MAGNDATA task), while a classifier that successfully separates the classes produces $F_1$ values substantially above this baseline.

\begin{figure*}[t]
\centering
\includegraphics[width=0.9\textwidth]{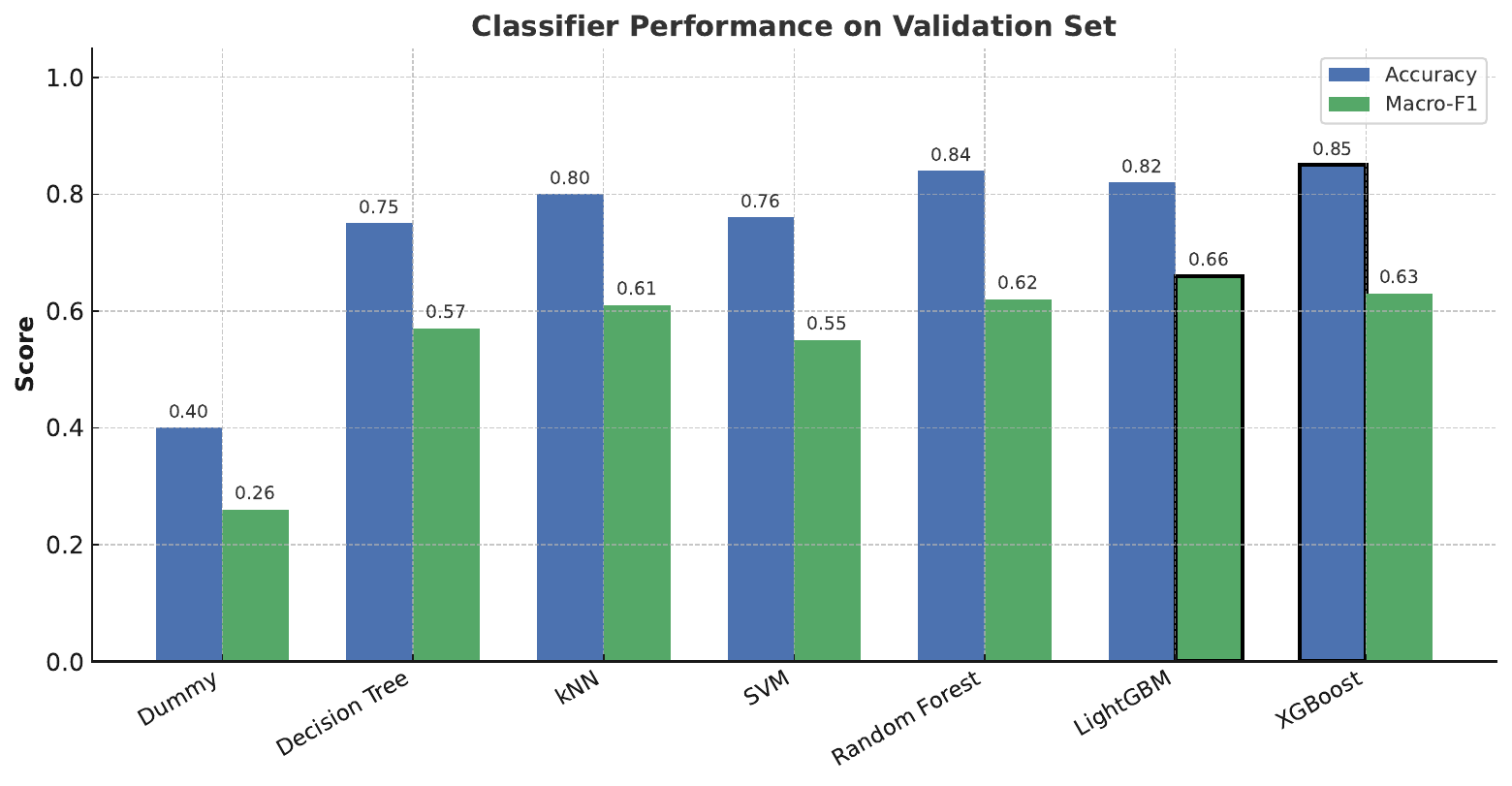}
\caption{Performance of different classifiers on the Materials Project validation set, measured by accuracy (blue) and macro-$F_1$ score (green). Simple baselines such as the dummy and decision tree classifiers perform significantly worse than ensemble methods, with Random Forest, LightGBM, and XGBoost achieving the highest accuracies (0.82–0.85) and balanced $F_1$ scores (0.62–0.66). Results shown here correspond to the four-class task for materials with at least one magnetic element, as discussed in the main text.}
\label{Performanceofclassifiers}
\end{figure*}

Because our MAGNDATA and MP datasets are class-imbalanced, we report the macro-averaged $F_1$ score, defined as
\begin{equation}
F_1^{\mathrm{macro}} = \frac{1}{k} \sum_{i=1}^{k} F_1^{(i)},
\label{eq:f1_macro}
\end{equation}
where $F_1^{(i)}$ is the $F_1$ score of class $i$ and $k$ is the number of classes. This gives each class equal weight regardless of its population, and therefore ensures that minority classes (AFM, FiM) contribute to the reported score as much as the majority classes (FM, NM), preventing a classifier that performs well only on majority classes from appearing deceptively strong overall. The $F_1$ score and its macro-averaged variant are standard metrics in the machine-learning literature~\cite{pedregosa2018scikitlearnmachinelearningpython,metrics1Powers2011,metrics2Sokolova2009} and are widely used for imbalanced classification tasks in materials science applications.  What qualifies as a high $F_1$ is task-dependent and varies with the inherent difficulty of the classification problem; for the magnetic-classification tasks considered in this work, a non-informative classifier (e.g. one that predicts labels at random in proportion to their class frequencies) produces $F_1^{macro}=1/k$ regardless of class imbalance (around 0.25 for our four-class MP task, see e.g. the Dummy Classifier in Table \ref{DummyClassifierTable}, and 0.5 for the binary MAGNDATA task).

\subsection*{ML Magnetic Ordering Class on The Materials Project}

As a first task, we benchmarked several supervised ML models, using \textit{scikit-learn} module \cite{pedregosa2018scikitlearnmachinelearningpython}, on the MP dataset: k-Nearest Neighbors, Decision Tree, LightGBM, Support Vector Machine, Random Forest, and XGBoost. To establish a true lower bound, we used \textit{scikit-learn}’s DummyClassifier with stratified strategy, which predicts labels at random, weighted by the class proportions in the training set. This baseline is intentionally feature-agnostic and its performance reflects only the dataset’s class imbalance independently from the features, and it is evaluated under the same 5-fold stratified CV, $20 \%$ validation, and $10\%$ test protocol as our real models (same splits, fixed \textit{random$\_$state}). The overall accuracy equals $\sum_kp_k^2$ where $p_k$ is the $k-$th class prior, and its per-class recall and precision are equal to $p_k$. This results in a per-class ${F_{1}}_ k = p_k$ and a macro-$F_1$ of $1/C$ where $C$ is the number of classes. Consequently, any model that does not outperform the DummayClassifier's accuracy ($\sum_kp_k^2$) and macro $F_1$ of $1/C$ captures no information beyond the classes priors. We check each of our ML models performance across the different CV folds to ensure that our results are reproducible and not artifacts of a particular data partition. For the MP dataset, performance of the baseline model is summarized in Table \ref{DummyClassifierTable}.

\begin{table}[]
\centering
\caption{Baseline performance of the stratified DummyClassifier on the Materials Project validation set.}
\label{DummyClassifierTable}
\begin{tabular}{lccc}
\hline\hline
Class & Precision & Recall & F1-score \\
\hline
AFM & 0.05 & 0.05 & 0.05 \\
FM  & 0.51 & 0.52 & 0.51 \\
FiM & 0.10 & 0.10 & 0.10 \\
NM  & 0.36 & 0.36 & 0.36 \\
\hline
Accuracy     &       &       & 0.40 \\
Macro avg    & 0.26  & 0.26  & 0.26 \\
Weighted avg & 0.40  & 0.40  & 0.40 \\
\hline\hline
\end{tabular}
\end{table}

As shown in Fig. \ref{Performanceofclassifiers}, all trained classifiers substantially outperformed the chance-level baseline model (accuracy$=0.40$, macro $F_1=0.26$) demonstrating that the features capture meaningful information about the underlying magnetic order. Simpler models such as a single decision tree (accuracy$=0.75$, macro $F_1=0.57$) and k-nearest neighbors (accuracy$=0.80$, macro $F_1=0.61$) improved considerably over the baseline but remained below the ensemble-based methods. Support vector machines achieved only a moderate performance (accuracy$=0.76$, macro $F_1=0.55$), likely due to the nonlinearity and the imbalance of our data. However, ensemble tree methods achieved the strongest results: XGBoost and Random Forest both reached accuracies above $84 \%$ with macro $F_1$ average scores around $0.63$ and $0.62$, respectively, while LightGBM delivered the highest macro $F_1$ score of $0.66$ at $0.82$ accuracy. Hyperparameter optimization for all classifiers was performed using RandomizedSearchCV or GridSearchCV with the search space and the selected hyperparameters reported in Table \ref{tab:hyperparameter_tuning_MP}. Given the balance between the overall correctness (captured by accuracy) and fairness across classes \textit{highlights LightGBM and XGBoost as the top performing models}, with LightGBM showing superior performance on underrepresented classes, and XGBoost providing the best aggregate predictive power. The strong performance demonstrated here using gradient-boosted tree ensembles in our study shows consistency with recent literature. For example, He et al. \cite{He_npjCompMat_2025} demonstrated the effectiveness of XGBoost in predicting the correct topological class of a given material, after being trained on the topological materials databases, with an accuracy of $85.2 \%$. Additionally, Ghosh et al. \cite{Ghosh_PRMaterials_2020} reported a Random Forest classifier for magnetic order of uranium compounds with a mean accuracy of $60.2\%$. These results collectively reinforces the powerfulness of the tree-based ML methods on learning complex material properties such as magnetism and topology using purely structural or derived features.

\begin{figure*}[t]
\centering
\subfloat[]{%
  \includegraphics[width=0.45\textwidth]{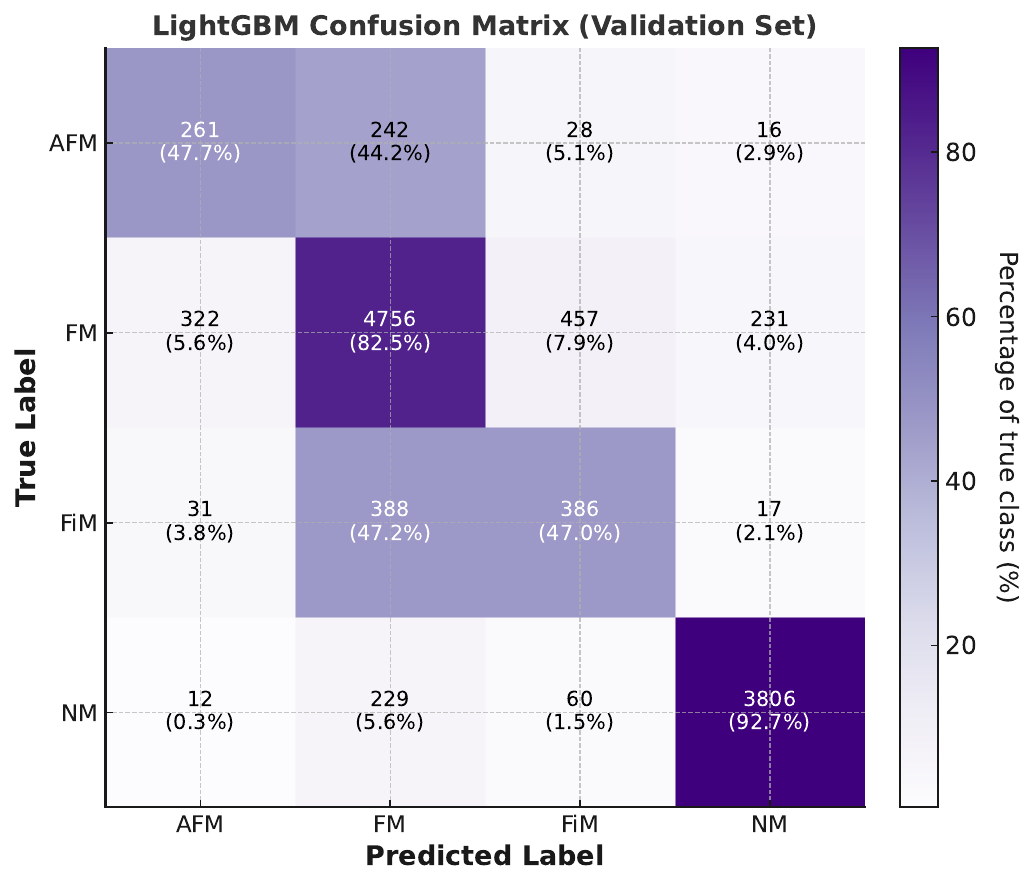}%
  \label{}}
\hfill
\subfloat[]{%
  \includegraphics[width=0.55\textwidth]{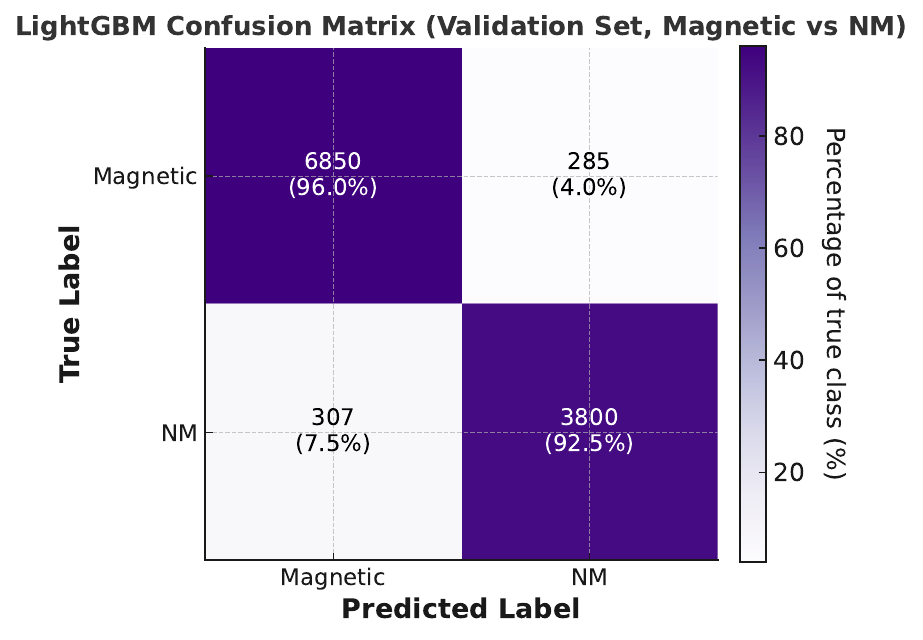}%
  \label{}}
\caption{Confusion matrices for the LightGBM classifier applied to the Materials Project dataset (restricted to compounds containing at least one magnetic element), shown under two labeling schemes: (a) four-class classification (FM, AFM, FiM, NM) and (b) binary classification (magnetic vs. nonmagnetic). Panel (a) corresponds to the classification performance reported in Fig. \ref{Performanceofclassifiers}.}
\label{LightGBMConfusionMatricesTwoClassAndFourClass}
\end{figure*}

As shown in the confusion matrix in Fig. \ref{LightGBMConfusionMatricesTwoClassAndFourClass} (a), the LightGBM classifier recognizes the FM and NM classes with a high fidelity ($F_1= 0.84$ and $0.93$, respectively), while the AFM and FiM classes remain more challenging ($F_1 \simeq 0.45$). This gap primarily arises from confusion between FM and FiM classes and the partial overlap between AFM and FM likely stemming from the fact that many materials in these classes share similar structural and compositional features, making them harder to distinguish using our current set of simple descriptors. In contrast, NM materials are readily identified with an $F_1$ score of $0.94$ showing that the model sees distinct features for this class. 

\begin{figure*}[t]
\hspace{-1.5cm}
\subfloat[]{%
  \includegraphics[width=0.45\textwidth]{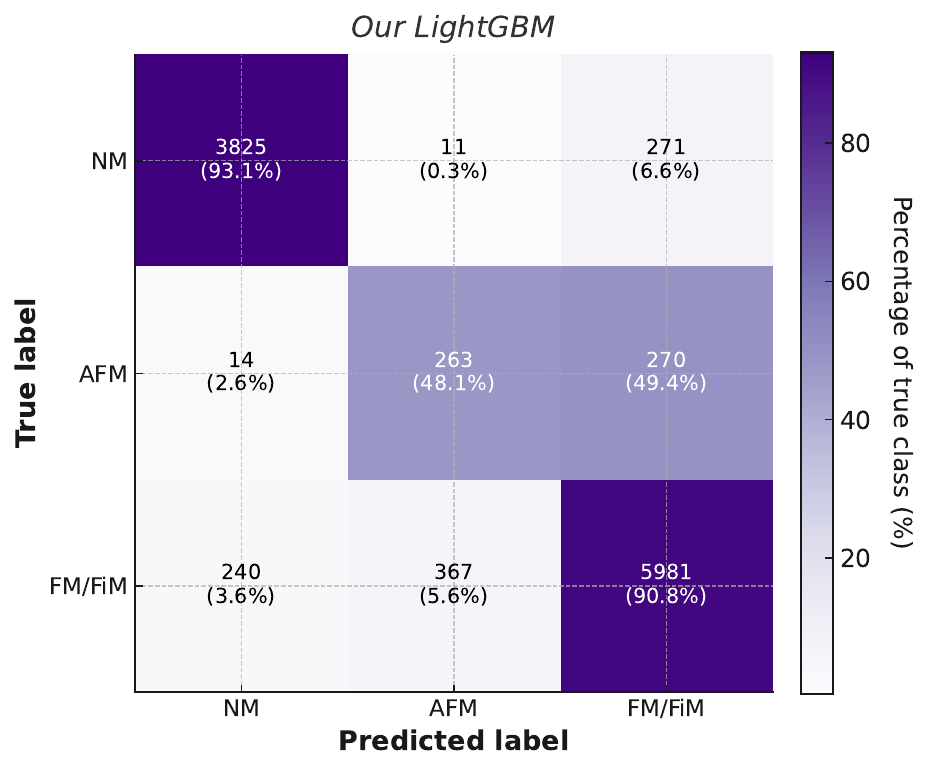}%
  \label{}}
\hspace{1.2cm}
\subfloat[]{%
  \includegraphics[width=0.45\textwidth]{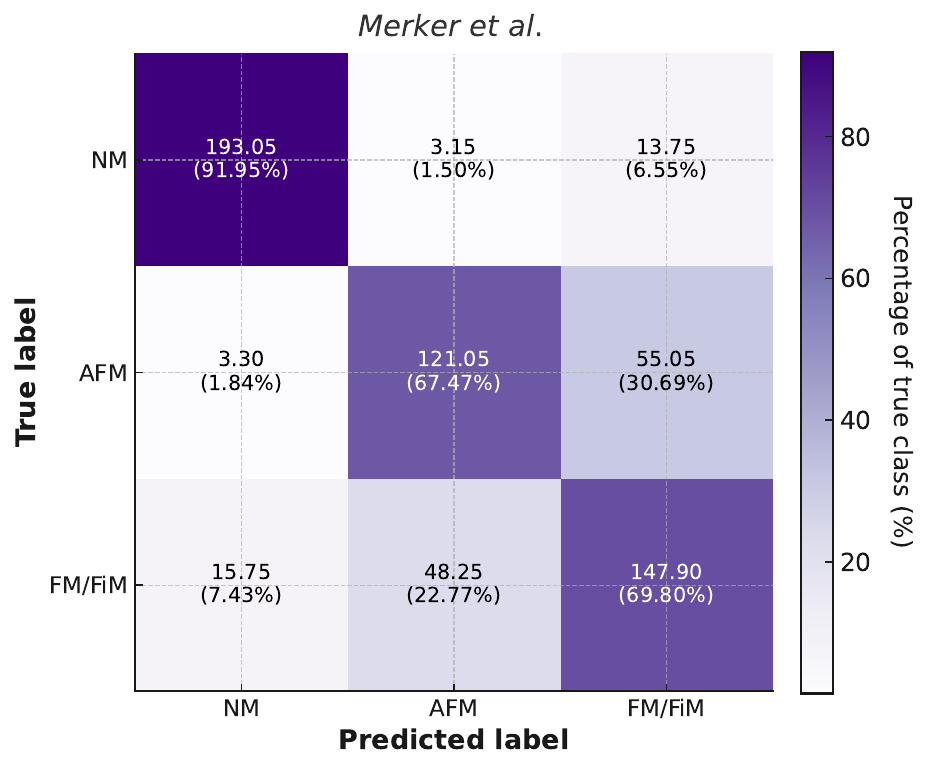}%
  \label{}}
\caption{Confusion matrices for: (a) the LightGBM classifier independently re-trained under a three-class labeling scheme (AFM, FM/FiM, NM) on the same Materials Project dataset, and (b) the result of Merker \emph{et al.}~\cite{Merker_iScience_2022}. The three-class scheme in panel (a) is adopted specifically for direct comparison with~\cite{Merker_iScience_2022}; small numerical differences relative to Fig. \ref{LightGBMConfusionMatricesTwoClassAndFourClass}(a), which uses the original four-class scheme, and to Fig. \ref{LightGBMConfusionMatricesTwoClassAndFourClass}(b), which uses the two-class scheme, reflect the independent model learning (e.g. different stratified splits and independently-tuned hyperparameters) under the two labeling schemes.}
\label{lightgbmVSmerkerPlusThreeClassesCase}
\end{figure*}

Now we compare the performance of LightGBM under different class groupings. When FM and FiM are merged into a single class FM$\_$FiM, the confusion matrix in Fig. \ref{lightgbmVSmerkerPlusThreeClassesCase} (a) shows an improvement: the majority of the FM$\_$FiM and NM materials are recovered with an accuracy over $90\%$, although AFM remains challenging to identify. Finally, in the binary setting of magnetic vs. nonmagnetic, shown in Fig. \ref{LightGBMConfusionMatricesTwoClassAndFourClass} (b), LightGBM achieves a remarkable separation with over $95\%$ correct classification of both classes. Again, this trend reflects the physical similarities of FM and FiM on one hand, and the scarcity of the AFM compounds on the MP database on the other hand.

To gauge our LightGBM relative to recent literature, we compare our $3-$class model (FM$\_$FiM, AFM, NM) with the results of Merker et al. \cite{Merker_iScience_2022}, as shown in Fig. \ref{lightgbmVSmerkerPlusThreeClassesCase}. It is critical to note that the two studies are working on different datasets: Merker et al \cite{Merker_iScience_2022} considers only compounds with calculations restricted to generalized gradient approximation and Hubbard interaction (GGA + U) \cite{Jain_APLMaterials_2013,Jain_HandbookMatModeling_2018}, resulting in a small dataset. In both, the recovery of the AFM class remains challenging, frequently interfering with the FM$\_$FiM class. Again, this reflects the fundamental challenge of distinguishing the AFM from the FM and FiM orders based solely on global descriptors. However, our LightGBM performs noticeably higher in both the FM$\_$FiM and NM classes, with classification rates of $90.8\%$ and $93.1\%$, respectively, compared to recovery rates of $69.80\%$ and $93.95\%$, respectively, while Merker et al. \cite{Merker_iScience_2022}'s model shows better performance on the AFM class with a recovery rate of $67.47\%$ compared to ours of $48.1\%$. \\

\begin{figure*}[t]
\hspace{-0.95cm}
\subfloat[Grouped feature importance]{%
  \includegraphics[width=0.49\textwidth]{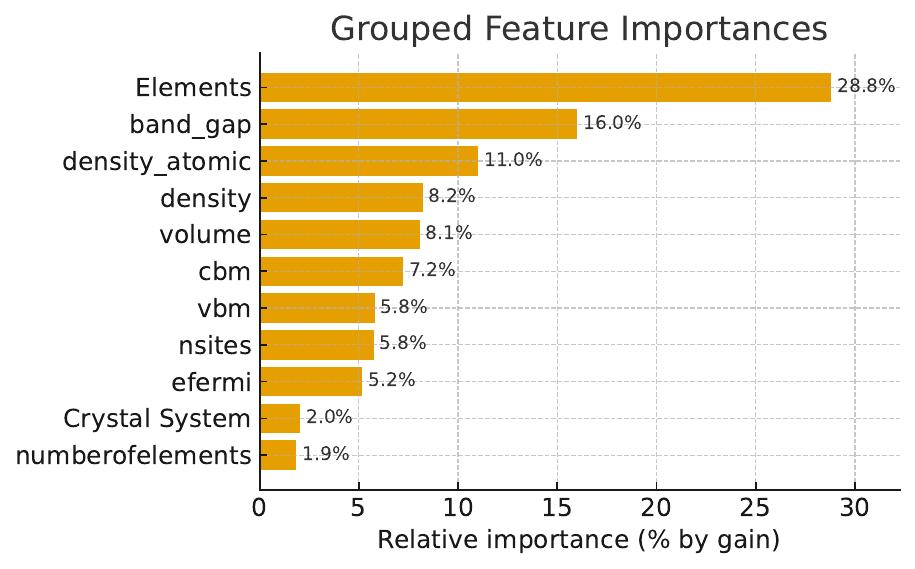}%
  \label{}}
\hspace{0.3cm}
\subfloat[Top 15 elemental contributions]{%
  \includegraphics[width=0.45\textwidth]{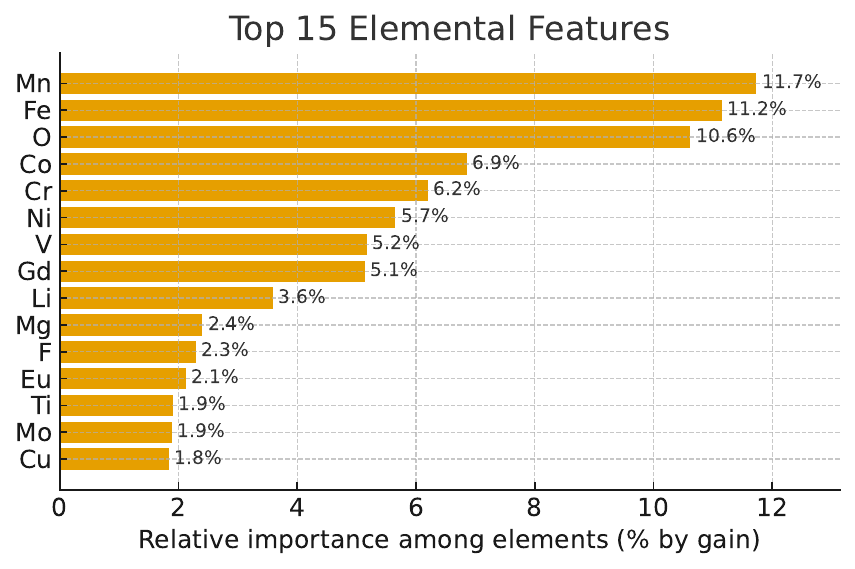}%
  \label{}}
\caption{Feature importance analysis based on LightGBM’s gain metric, which aggregates the reduction in the loss function contributed by each feature across all decision-tree splits. (a) Grouped feature importances, highlighting the dominant role of compositional descriptors (elements), electronic band gap, and atomic density. (b) Top 15 elemental contributions, showing Mn, Fe, and O as the most influential elements for classification performance.}
\label{LightGBMfeatureimportance}
\end{figure*}

\paragraph*{Feature importance.}The feature importance analysis was performed using LightGBM's gain-based improtance measure, which sums the reduction in the loss function contributed by each given feature across all decision-tree splits. The importance of a feature $f$ is defined as:
$$I_{gain}(f)=\sum_{s\in S(f)}\Delta L(s)$$
where $S(f)$ is the set of all splits that employ the feature $f$, and $\Delta L(s)$ is the improvement achieved for the loss function through this split. Therefore, this measure provides a distinction for features that have the strongest predictive power rather than those that are not used frequently. The results show that elemental composition dominates, with around one third of the total model gain (see Fig. \ref{LightGBMfeatureimportance} (a)). Among the numerical features, the electronic band gap shows high importance, possibly due to its central role in determining the electronic ground states and their connection to magnetic order. Structural-related descriptors such as the atomic and overall density and volume, along with other electronic features such as CBM, VBM, and Fermi energy, also rank highly. This underscores the fundamental interplay between structural packing, electronic filling, and magnetic order. Crystal system contributes only a few percent, suggesting that global crystal type does not efficiently distinguish magnetic orders alone without extra features. This originates from the fact that the crystal structure (cubic, tetragonal, orthorhombic, etc.) is a coarse-grained symmetry classification that captures the metric of the unit cell and constrains the allowed point-group symmetries of the unit cell but discards information about the local coordination of magnetic ions and the bond-angle distributions that govern superexchange (per the Goodenough-Kanamori-Anderson rules~\cite{Goodenough1963}). Two compounds in the same crystal system can therefore exhibit very different magnetic ground states, and compounds in different crystal systems can share a common magnetic mechanism. Individually examined, the elemental contribution is dominated by $3d$ transition metals such as Mn, Fe, Co, Cr, and Ni, with oxygen also highly performing (see Fig. \ref{LightGBMfeatureimportance} (b)), possibly due to its role in mediating superexchange pathways in many compounds in the database. This physically reflects the fact that transition-metal chemistry and their surrounding anion environment being critical to the magnetic interactions and therefore the emergent magnetic order.

However, it is important to distinguish between two qualitatively different ways in which a feature can be informative to a tree-ensemble classifier. A feature may be predictive because it directly encodes a physical quantity that influences the target — for example, the band gap, Fermi level, unit-cell volume, or site count, each of which reflects the underlying electronic structure, hybridization strength, and density of magnetic centers relevant to magnetic ordering. Alternatively, a feature may be predictive only as a statistical proxy for a chemical or structural family that is over- or under-represented in the training data, with no direct physical mechanism linking it to the target. Feature-importance rankings from tree ensembles conflate these two modes and cannot, by themselves, distinguish them. We therefore caution the reader against inferring a physical role from such rankings without an independent mechanism-based justification which is beyond the scope of this work.

Evaluating the performance of LightGBM over the final held-out test set ($10\%$), shown in Table. \ref{LightGBMTestAndValidation}, results in a classification accuracy of $83\%$ and a macro $F_1$ average score of $69\%$, which closely matches the performance over the validation set of $82\%$ accuracy and $66\%$ macro $F_1$ average score. The consistency of the classifier's performance over the two subsets indicates that the model generalizes well with no evidence of overfitting to the training distribution. Class-level metrics are also preserved: NM and FM classes are highly recovered, while AFM and FiM classes achieve lower $F_1$ scores, but their recall and precision remain comparable across the two subsets. Stability of the performance is confirmed as well during the five-fold cross validation strategy adopted at the training stage. This stability supports the robustness of the LightGBM classifier and its applicability to unseen compounds.
\begin{table*}
  \hspace{-1cm}
  \caption{LightGBM performance by class on the validation and test sets. Values reported to two decimals.}
  \label{LightGBMTestAndValidation}
  \renewcommand{\arraystretch}{1.15}
  \setlength{\tabcolsep}{6pt}
  \begin{tabular}{lccc|ccc}
    \toprule
    & \multicolumn{3}{c}{\textbf{Validation}} & \multicolumn{3}{c}{\textbf{Test}} \\
    \cmidrule(lr){2-4}\cmidrule(lr){5-7}
    \textbf{Class} & \textbf{Prec.} & \textbf{Rec.} & \textbf{F1} & \textbf{Prec.} & \textbf{Rec.} & \textbf{F1} \\
    \midrule
    AFM & 0.42 & 0.48 & 0.45 & 0.46 & 0.54 & 0.50 \\
    FM  & 0.85 & 0.82 & 0.84 & 0.85 & 0.83 & 0.84 \\
    FiM & 0.41 & 0.47 & 0.44 & 0.45 & 0.50 & 0.47 \\
    NM  & 0.94 & 0.93 & 0.93 & 0.93 & 0.93 & 0.93 \\
    \midrule
    \textbf{Overall acc.} & \multicolumn{3}{c}{0.82} & \multicolumn{3}{c}{0.83} \\
    \textbf{Macro avg}    & 0.65 & 0.67 & 0.66 & 0.68 & 0.70 & 0.69 \\
    \textbf{Weighted avg} & 0.83 & 0.82 & 0.82 & 0.83 & 0.83 & 0.83 \\
    \bottomrule
  \end{tabular}
\end{table*}
\subsection*{ML Propagation Vector on MAGNDATA}
\begin{figure*}[t]
\centering
\subfloat[]{%
  \includegraphics[width=0.5\textwidth]{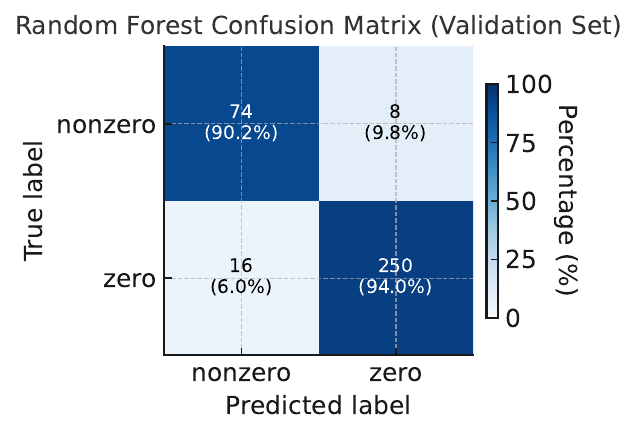}%
  \label{}}
\hfill
\subfloat[]{%
  \includegraphics[width=0.45\textwidth]{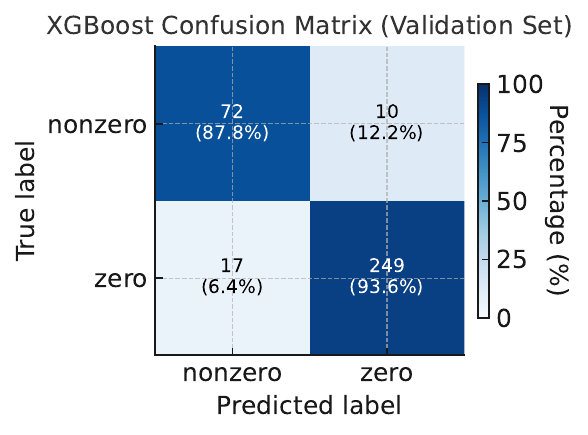}%
  \label{}}
\caption{Confusion matrices for propagation-vector classifiers on MAGNDATA: (a) Random Forest and (b) XGBoost.}
\label{PropagationVectorConfusionMatrixRFXGB}
\end{figure*}

For the second classification task, we focus on the second dataset formed by supplementing the materials extracted from MAGNDATA with features from the MP database, as described earlier. We train two classifiers: Random Forest and XGBoost on learning the binary propagation vector (zero vs. nonzero). Even though this binary information is simpler than distinguishing different magnetic orders, it is still physically meaningful since a nonzero propagation vector implies a modulated or complex magnetic structure, which shall also help us in the next section to diagnose the FM bias on the MP database. We adopt the same data splitting strategy as discussed earlier with the same five-fold cross validation strategy at the training step. Both models achieved strong performance on the validation set, with XGBoost reaching an overall accuracy of $92\%$ and a macro $F_1$ average score of $90\%$, while Random Forest slightly outperforms it with an accuracy of $93\%$ and $91\%$ macro $F_1$ average score. Hyperparameter optimization is reported in Table \ref{tab:hyperparameter_tuning_magndata}. In both models, the zero propagation vector class was classified with a very high recall and precision (around $94\%$), while the nonzero case showed slightly lower but still robust with a recall of $88-90\%$ and a precision of $81-82\%$. As shown in the confusion matrix in Fig. \ref{PropagationVectorConfusionMatrixRFXGB}, the majority of the error stems from mislabeling a small fraction of zero-$k$ cases as nonzero-$k$ materials, while the nonzero-$k$ structures are recovered with higher reliability. 

\begin{figure*}[]
\centering
\subfloat[]{%
  \includegraphics[width=0.47\textwidth]{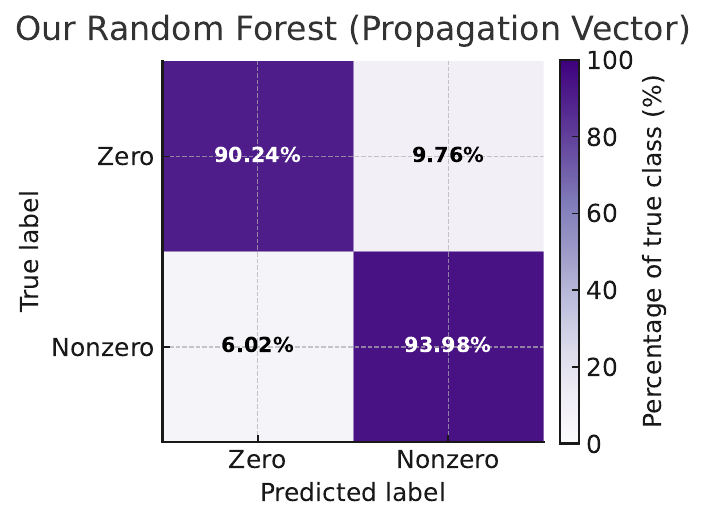}%
  \label{}}
\hfill
\subfloat[]{%
  \includegraphics[width=0.45\textwidth]{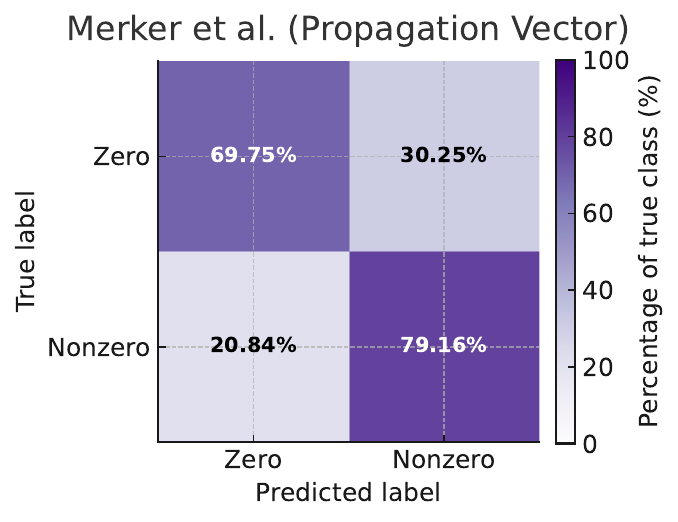}%
  \label{}}
\caption{Comparison of propagation vector classification results (zero vs.\ nonzero) shown as row-normalized confusion matrices with percentage values for (a) our Random Forest model trained on MAGNDATA and (b) results from Merker \textit{et al.} \cite{Merker_iScience_2022}.}
\label{RandomForestVSMerkerkVector}
\end{figure*}

\begin{figure}[h]
    \centering
    \includegraphics[width=1\linewidth]{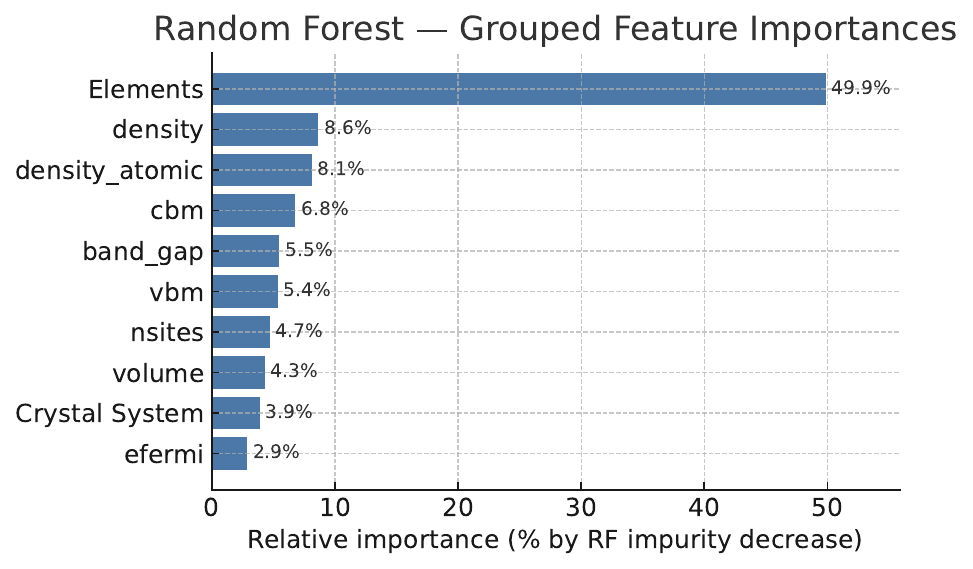}
    \caption{Random Forest feature importance analysis for propagation vector classification on the MAGNDATA dataset. Grouped feature importances, showing that compositional descriptors (elements) dominate, followed by density-related and electronic features such as atomic density, band gap, and conduction/valence band edges.}
    \label{PropagationFeatureImportance}
\end{figure}

Highlighting Random Forest as the best propagation vector classifier (even though it slightly outperforms XGBoost), we utilize its feature importance analysis (see Fig. \ref{PropagationFeatureImportance}) to show that the chemical composition of a given compound provides the largest contribution, followed by structural descriptors such as density and atomic density, and electronic descriptors such as band gap, CBM, and VBM. The crystal system still shows lower predictive power. Finally, the model generalizes well to unseen materials, as confirmed in the five-fold cross-validation process, and its performance over the final test set. To benchmark the strong performance of our Random Forest, we compare it to the results of \cite{Merker_iScience_2022}, as shown in Fig. \ref{RandomForestVSMerkerkVector}. A similar analysis holds for the XGBoost classifier.

\begin{table*}[t]
\caption{Expanded literature audit of candidate non-ferromagnetic compounds from Supplementary Table S1 (intersection set between the XG Boost and Random Forest MAGNDATA classifiers) with references confirming their non-FM ordering.}
\label{tab:s1_lit_audit}
\scriptsize
\begin{ruledtabular}
\begin{tabular}{lll}
Material & Reference & \;\;\;\;\;\;\;\;\;\;\;\;\;\;\;\;\;\;\;\;\;\;\;\;\;\;\;\; \;\;\;\;\;\;\;\;\;\;\;\;\;\;\;\;\;Comments \\
Ba$_2$GdNbO$_6$ &
Ref.~\cite{berardi2025bagnbo} &
\parbox[t]{8.4cm}{Predominantly paramagnetic down to 1.8 K.} \\
Ba$_2$GdSbO$_6$ &
Ref.~\cite{Ba2GdSbO6koskelo2022freespindominatedmagnetocaloriceffect} &
\parbox[t]{8.4cm}{Heat capacity measurements indicate a lack of magnetic ordering down to 0.4 K.} \\
Ba$_2$GdTaO$_6$ &
Ref.~\cite{doi2001ba2lntao6} &
\parbox[t]{8.4cm}{Magnetic susceptibility measurements shows paramagnetism down to 5 K.} \\
CaV$_4$O$_9$ &
Ref.~\cite{sano1996cav4o9} &
\parbox[t]{8.4cm}{Two-dimensional spin-gap antiferromagnet.} \\
Co$_2$PO$_4$F &
Ref.~\cite{leblanc1997co2po4f} &
\parbox[t]{8.4cm}{Antiferromagnetic with $T_N \approx 14$ K and negative Curie--Weiss temperature.} \\
CsCr(MoO$_4$)$_2$ &
Ref.~\cite{springermaterials_cscrmoo4} &
\parbox[t]{8.4cm}{Springer Materials classifies it as AFM.} \\
FeCl$_3$ &
Ref.~\cite{Levinsky2025_FeCl3_MaterAdv} &
\parbox[t]{8.4cm}{Single-crystal neutron diffraction and SANS on a non-centrosymmetric polymorph reveal an antiferromagnetic transition at $T_N = 8.6$ K with propagation vector $\mathbf{k}=(1/2,0,1/3)$.} \\
$\alpha$-Gd$_2$S$_3$ &
Ref.~\cite{Ebisu2004_Gd2S3_JPCS} &
\parbox[t]{8.4cm}{Orthorhombic (\textit{Pnma}) Gd$^{3+}$ sulfide; magnetic susceptibility of single-crystal samples shows long-range antiferromagnetic ordering below $\sim$10 K.} \\
HCrO$_2$ &
Ref.~\cite{somesh2021hcro2} &
\parbox[t]{8.4cm}{Triangular-lattice antiferromagnet} \\
KCrS$_2$ &
Ref.~\cite{vanlaar1973kcrs2} &
\parbox[t]{8.4cm}{Magnetic structure consists of ferromagnetic sheets coupled antiferromagnetically.} \\
KFe(MoO$_4$)$_2$ &
Ref.~\cite{smirnov2010kfemoo4} &
\parbox[t]{8.4cm}{Stacked triangular antiferromagnet; $T_N \approx 2.5$ K; coexistence of one collinear AFM subsystem and one spiral subsystem.} \\
LaCrTeO$_6$ &
Ref.~\cite{springermaterials_lacrteo6} &
\parbox[t]{8.4cm}{Springer Materials classifies LaCrTeO$_6$ as AFM.} \\
Mn$_2$O$_3$ &
Ref.~\cite{Cong2018_Mn2O3_NatCommun} &
\parbox[t]{8.4cm}{Mn$_2$O$_3$ exhibits two intricate antiferromagnetic structures below $\sim$100 K (neutron powder diffraction).} \\
Mn$_5$O$_8$ &
Ref.~\cite{Punnoose2001_Mn5O8_IEEETMAG} &
\parbox[t]{8.4cm}{Magnetic susceptibility and ESR show antiferromagnetic ordering with $T_N \approx 128$ K.} \\
MnSe &
Ref.~\cite{Man2022_MnSe_PRB} &
\parbox[t]{8.4cm}{The high-pressure orthorhombic phase of MnSe shows AFM behavior.} \\
NaCrS$_2$ &
Ref.~\cite{blazey1969nacrs2} &
\parbox[t]{8.4cm}{Antiferromagnetic ordering observed below about 18 K.} \\
Pr$_2$NiO$_4$ &
Ref.~\cite{fernandezdiaz1991pr2nio4} &
\parbox[t]{8.4cm}{Ni sublattice becomes three-dimensionally antiferromagnetically ordered at $T_N \approx 325$ K with propagation vector $k=[100]$.} \\
Rb$_2$V$_3$O$_8$ &
Ref.~\cite{liu1995rb2v3o8} &
\parbox[t]{8.4cm}{Fresnoite-type vanadate with antiferromagnetic interactions below $\sim 10$ K.} \\
RbFe(MoO$_4$)$_2$ &
Ref.~\cite{svistov2003rbfemoo4} &
\parbox[t]{8.4cm}{Quasi-two-dimensional triangular-lattice antiferromagnet with noncollinear $120^\circ$ order below $T_N \approx 3.8$ K.} \\
VBr$_3$ &
Ref.~\cite{Gu2024_VBr3_PRB} &
\parbox[t]{8.4cm}{Van der Waals honeycomb antiferromagnet; neutron diffraction reveals magnetic Bragg peaks below $T_N = 26.5$ K with $\mathbf{k}=(0,1/2,1)$, corresponding to a zigzag AFM structure.} \\
VF$_3$ &
Ref.~\cite{Gossard1972_VF3_AIPCP} &
\parbox[t]{8.4cm}{NMR/susceptibility studies identify an antiferromagnetic transition at with $T_N \approx 19.4$ K.} \\
YMn$_2$O$_5$ &
Ref.~\cite{radaelli2008ymn2o5} &
\parbox[t]{8.4cm}{Antiferromagnetic domains observed directly by neutron spherical polarimetry.} \\
\end{tabular}
\end{ruledtabular}
\end{table*}

\subsection*{MAGNDATA ML Classifiers Applied onto the Materials Project Database}
In our final task, we employ our propagation vector classifiers trained on MAGNDATA to the MP database in order to assign each compound a binary propagation vector label. This enables a large-scale prediction of the binary $k-$vector, a goal that is not achievable on the MP database alone. Through the predictions of this classifier, our goal is to identify compounds that are labeled as ferromagnets on the Materials Project database, yet predicted to have a nonzero $k-$vector through the MAGNDATA-trained-on classifiers. Before that, we filter out MAGNDATA materials that are available on the MP database to avoid data leakage as discussed in section \ref{sec:data collecting}. We also use the MP dataset without filtering out materials that do not have at least one magnetic ion in order to test the MAGNDATA classifiers capability of capturing such cases. Applying the $k-$vector Random Forest classifier, we identify $16,973$ materials labeled as ferromagnets on the Materials Project database predicted to have a non-zero propagation vector. Due to the existence of more than one MP profile for some materials on the MP database, among these are $10,354$ unique materials (\textit{see Supplementary Table 2}). Repeating this for the $k-$vector XGBoost, $14,064$ (among them are $9,383$ unique compunds, \textit{see Supplementary Table 3}) MP-labeled FMs are predicted to have a nonzero propagation vector. Given the strong performance of the $k-$vector classifiers on MAGNDATA, which is a realistic database compared to the DFT-based Materials Project database, we rely on the mentioned \textit{contradiction} above $-$ having a predicted nonzero $k-$vector and a FM label on the MP database $-$ to predict that these materials are candidate wrongly-labeled FMs on the MP database, therefore partially address its bias towards the FM class. As a practice of having more confidence in the final set of potentially mislabeled MP-ferromagnets, we choose the intersecting set of compounds of $7,843$ (\textit{see Supplementary Table 1}) between the two $k-$vector classifiers. A few examples of our tabulated candidate MP-mislabeled FM materials include FeCl$_3$ \cite{Levinsky2025_FeCl3_MaterAdv}, VBr$_3$ \cite{Gu2024_VBr3_PRB}, VF$_3$ \cite{Gossard1972_VF3_AIPCP}, $\alpha$-Gd$_2$S$_3$ \cite{Ebisu2004_Gd2S3_JPCS}, MnSe \cite{Man2022_MnSe_PRB}, Mn$_5$O$_8$ \cite{Punnoose2001_Mn5O8_IEEETMAG}, and Mn$_2$O$_3$ \cite{Cong2018_Mn2O3_NatCommun} (see Table \ref{tab:s1_lit_audit} for more materials), where the associated references validate our predictions of the existence of non-ferromagnetic orders.

\paragraph*{Estimating the false discovery rate.}Extending the naive propagation of the individual-classifier precisions (82\% for RF and 81\% for XGBoost) over the MAGNDATA validation dataset to the 7,843-compound intersection set gives false discovery rates of $\mathrm{FDR}_{\mathrm{RF}} = 1 - 0.82 = 18\%$ and $\mathrm{FDR}_{\mathrm{XGB}} = 1 - 0.81 = 19\%$, corresponding to approximately 1,400--1,500 expected false positives. This estimate, however, is a loose upper bound: it assumes that individual-classifier errors carry over unchanged to the intersection of two classifiers trained with different inductive biases, which overstates the true intersection-set FDR.

To quantify the false-discovery rate empirically of the intersection set (Supplementary Table S1), we drew a uniformly random sample of 100 compounds from Supplementary Table S1 and conducted a systematic literature audit of each (see Table \ref{tab:expanded_audit}). Of the 100 sampled compounds, zero were confirmed to be ferromagnetic. Four are confirmed non-FM by direct experiment on the specific compound (BaY$_2$NiO$_5$, Eu$_2$MgSi$_2$O$_7$, Mn$_2$ZnO$_4$, TmMn$_2$O$_5$), 22 are likely non-FM based on established behavior of their structural or chemical family, and 16 cannot be ferromagnetic because they contain no magnetic ions --- their MP FM label is a pure DFT labeling artifact. Five compounds reside in chemical regimes where FM ordering could plausibly occur --- two mixed-valence manganites supporting Zener double exchange (Ca$_2$HoMn$_4$BiO$_{12}$, SrPr$_2$Mn$_3$O$_9$), one Cr-chalcogenide spinel (Cr$_4$InAgSe$_8$) by analogy with the FM insulator family CdCr$_2$Se$_4$/HgCr$_2$Se$_4$, one rare-earth intermetallic with RKKY coupling (Gd(Tl$_3$Te$_2$)$_3$), and one DFT-predicted FM insulator (NaCr$_3$O$_8$) --- and are conservatively classified as potential false positives. The remaining 53 compounds have no published experimental magnetic data with which to compare directly. Even under the most conservative accounting --- treating all five potential-FM candidates as genuine false positives --- the empirical FDR on this random sample is bounded at $\lesssim 5\%$, less than a third of the naive $18\text{--}19\%$ ceiling obtained by propagating individual-classifier precisions. This validates the premise that taking the intersection of two classifiers with different inductive biases substantially suppresses error rates below either model alone.

\paragraph*{Removing noisy entries from the candidate material datasets.}Finally, direct screening of the full intersection set shows that 1,003 of the 7,843 candidates ($12.8\%$) contain no magnetic ions at all, confirming that a substantial fraction of MP's FM labels are pure DFT labeling artifacts — consistent and close to the $16\%$ "no magnetic ion" rate observed in our random-sample audit (Table \ref{tab:expanded_audit}). Candidate non-FM materials with at least one magnetic atom predicted by each classifier have a similar fraction of FM artifacts and are also listed in the \textit{Supplementary Tables 4-6}, where the number of XGBoost flagged candidates reduce from 9,383 to 8,189 ($12.7\% $ are DFT artifacts), Random Forest from 10,354 to 9,055 ($12.5\% $ are DFT artifacts), and their intersection set reduces from 7,843 to 6,840. Although our classifiers correctly identify these compounds as having nonzero propagation vector (i.e., not simple FM), flagging such compounds as candidates for being FM-mislabeled is trivially correct. Nonetheless, we retain the D-category entries are retained in Table \ref{tab:expanded_audit} because they independently show that the classifiers are still capable of identifying these compounds.

\paragraph*{Effect of additional extra physically motivated descriptors.} Having established the performance of the classifiers using our minimal set of descriptors, we now relax the restriction described in Section~II and re-train the classifiers with two additional physically motivated compositional descriptors: the composition-weighted mean Pauling electronegativity and the composition-weighted mean atomic dipole polarizability. These descriptors are related to screened Coulomb interactions and have been used as predictors in a recent machine-learning study of magnetic ordering~\cite{Merker_iScience_2022}. Per-element electronegativities were obtained from \texttt{pymatgen}~\cite{Ong2013pymatgen}, and per-element polarizabilities from the \texttt{mendeleev} Python package~\cite{mendeleev}, which sources its values from the CRC Handbook of Chemistry and Physics \cite{CRC_Handbook} and from Ref.~\cite{Schwerdtfeger2019}.

The effect on the performance of our best classifiers is summarized in Tables~\ref{tab:descriptor_ablation} and \ref{tab:descriptor_ablation_RF_kvector}. The cross-validated $F_1^{\mathrm{macro}}$ improves by approximately $1\%$ on the four-class MP task across the different classifiers (e.g. from $0.658$ to $0.665$ for LightGBM) and by approximately $1-2\%$ on the binary MAGNDATA propagation-vector task (e.g. from $0.883$ to $0.906$ for Random Forest). The improvement is more concentrated on the minority classes, as expected when more discriminative features benefit the harder-to-distinguish categories more. The modest overall increase in the classifiers efficiency indicates that the tree-ensemble classifiers could already implicitly capture much of the compositional-averaging information that explicit means of unseen descriptors encode. A more substantial gain would likely require local structural features --- coordination environments, magnetic-site connectivity, or other descriptors that global compositional means cannot encode.

\begin{table}[h]
\caption{Effect of adding composition-weighted mean Pauling electronegativity ($\bar{X}$) and atomic dipole polarizability ($\bar{\alpha}$) on the Random Forest classifier performance for the binary propagation-vector classification task on MAGNDATA. Per-class precision, recall, and $F_1$ are reported on the validation set. CV $F_1$-macro is the mean over 6 stratified folds on the training set.}
\label{tab:descriptor_ablation_RF_kvector}
\small
\begin{ruledtabular}
\begin{tabular}{lcccccc}
 & \multicolumn{3}{c}{Without $\bar{X}, \bar{\alpha}$} & \multicolumn{3}{c}{With $\bar{X}, \bar{\alpha}$} \\
\cline{2-4}\cline{5-7}
Class & Prec. & Rec. & $F_1$ & Prec. & Rec. & $F_1$ \\
\hline
nonzero       & 0.82 & 0.90 & 0.86 & 0.84 & 0.90 & 0.87 \\
zero          & 0.97 & 0.94 & 0.95 & 0.97 & 0.95 & 0.96 \\
\hline
Accuracy      & \multicolumn{3}{c}{0.93} & \multicolumn{3}{c}{0.94} \\
Macro avg     & 0.90 & 0.92 & 0.91 & 0.91 & 0.92 & 0.91 \\
Weighted avg  & 0.93 & 0.93 & 0.93 & 0.94 & 0.94 & 0.94 \\
\hline
Mean CV $F_1^{\mathrm{macro}}$ & \multicolumn{3}{c}{0.883} & \multicolumn{3}{c}{0.906} \\
\end{tabular}
\end{ruledtabular}
\end{table}

\begin{table}[h]
\caption{Effect of adding composition-weighted mean Pauling electronegativity ($\bar{X}$) and atomic dipole polarizability ($\bar{\alpha}$) on the LightGBM classifier performance for the four-class MP magnetic-ordering task. Per-class precision, recall, and $F_1$ are reported on the validation set. CV $F_1$-macro is the mean over $5$ stratified folds on the training set.}
\label{tab:descriptor_ablation}
\small
\begin{ruledtabular}
\begin{tabular}{lcccccc}
 & \multicolumn{3}{c}{Without $\bar{X}, \bar{\alpha}$} & \multicolumn{3}{c}{With $\bar{X}, \bar{\alpha}$} \\
\cline{2-4}\cline{5-7}
Class & Prec. & Rec. & $F_1$ & Prec. & Rec. & $F_1$ \\
\hline
AFM           & 0.42 & 0.48 & 0.45 & 0.45 & 0.46 & 0.46 \\
FM            & 0.85 & 0.82 & 0.84 & 0.85 & 0.85 & 0.85 \\
FiM           & 0.41 & 0.47 & 0.44 & 0.45 & 0.49 & 0.47 \\
NM            & 0.94 & 0.93 & 0.93 & 0.94 & 0.93 & 0.94 \\
\hline
Accuracy      & \multicolumn{3}{c}{0.82} & \multicolumn{3}{c}{0.83} \\
Macro avg     & 0.65 & 0.67 & 0.66 & 0.67 & 0.68 & 0.68 \\
Weighted avg  & 0.83 & 0.82 & 0.82 & 0.84 & 0.83 & 0.83 \\
\hline
Mean CV $F_1^{\mathrm{macro}}$ & \multicolumn{3}{c}{0.658} & \multicolumn{3}{c}{0.665} \\
\end{tabular}
\end{ruledtabular}
\end{table}

\section{Conclusion}
In conclusion, a central outcome of this work is the identification of a systematic ferromagnetic bias in the Materials Project database, revealed through our propagation-vector classifiers trained on experimentally validated MAGNDATA entries. While Materials Project magnetic labels frequently default to FM orderings stemming from the use of ferromagnetic initialization in automated DFT workflows, even when the true ground state is antiferromagnetic, ferrimagnetic, or complex, our models consistently flagged these discrepancies for more than $6,840$ candidate compounds, with the empirical false-discovery rate estimated by a random-sample validation to be $\lesssim 5\%$. Specifically, the Random Forest and XGBoost classifiers we developed achieved accuracies of over $92\%$ (with $F_1$-macro scores above $90\%$) in distinguishing zero from nonzero propagation vector structures. Importantly, our approach demonstrates how machine learning can serve not only as a predictive tool but also as a diagnostic layer to expose and correct database-level artifacts, offering a scalable pathway to improve the reliability of high-throughput materials databases.

\paragraph*{Limitations.} However, our approach has several limitations that should be kept in mind. First, the class distribution in MAGNDATA is strongly unbalanced: antiferromagnetic entries dominate the database, with ferromagnetic, and ferrimagnetic compounds much more sparsely represented. As a result, a classifier trained directly on MAGNDATA cannot reliably learn to distinguish all four magnetic classes simultaneously, and this is the main reason we opted for a binary propagation-vector classifier (zero versus nonzero) rather than a four-class magnetic-order classifier on MAGNDATA. Second, although the tree-based architectures used in this work are efficient in classification tasks \cite{treemodelsgrinsztajn2022treebasedmodelsoutperformdeep}, they are still insufficient to predict the full three-dimensional magnetic structure of a given compound. Our classifiers answer coarse-grained binary or four-way questions, but reconstructing the complete spin configuration — including the direction of the ordered moments, the detailed periodicity of modulated phases, and non-collinear arrangements — requires more expressive architectures such as graph neural networks that operate directly on atomic coordinates and local environments. Third, our feature set is deliberately minimal, consisting only of elemental composition, and a handful of global electronic and structural quantities available from the Materials Project API. Richer descriptors — including coordination environments, bond-length distributions, magnetic-site connectivity graphs, and explicit spin-orbit or Hubbard
U estimates — would likely improve both accuracy and physical interpretability, and could allow the classifiers to resolve finer distinctions such as collinear versus spiral antiferromagnetism that are currently merged into the single nonzero propagation vector class.

\paragraph*{Future Directions.} Now we suggest several concrete directions as natural next steps to build on this work. Perhaps the most immediate is to move beyond the tree-ensemble, global-descriptor approach and adopt architectures that operate directly on atomic structure, such as crystal graph convolutional networks and Euclidean equivariant neural networks~\cite{Merker_iScience_2022, Choudhary2024}, which preserve crystallographic symmetry by construction and can in principle predict the full three-dimensional magnetic structure rather than the binary summary we consider here. More recent spin-dependent and time-reversal equivariant networks~\cite{Yu2024SpinGNN} go further and explicitly incorporate the spin degree of freedom into the network architecture, allowing direct prediction of non-collinear configurations and spin-lattice coupling. Additionally, some materials exhibit more than one magnetic structure depending at the temperature, therefore incorporating temperature effects on the formation of the magnetic order is a higher level future task. A second direction is to enlarge and rebalance the training data: the recently compiled Northeast Materials Database (NEMAD), which contains about 67{,}500 experimental-based reported magnetic entries extracted from the literature using large language models~\cite{Itani2025NEMAD}, offers a substantially larger and more class-balanced training set than MAGNDATA, and incorporating it would partially address the class-imbalance limitation we discuss above. Where balancing the database itself is not possible, data-level interventions such as SMOTE-style oversampling \cite{SMOTEChawla_2002} and loss-level cost-sensitive learning \cite{khan2017costsensitivelearningdeep} are both straightforward to integrate into the existing pipeline. A third direction is to close the loop between the machine learning screen and the underlying \emph{ab initio} calculations: our flagged candidate list can serve as input to an automated DFT+$U$ workflow with explicit AFM enumeration in the spirit of Horton \emph{et al.}~\cite{Horton2019MP}, which would convert our heuristic flags into concrete proposed corrections to the MP labels and, through active-learning-style feedback~\cite{Long2022PNAS}, progressively refine the classifiers themselves. Fourth, targeted experimental follow-up on high-interest compounds from the flagged list --- particularly those whose chemistry is unusual enough that no obvious family-level analog exists --- remains the most decisive test of the approach, and the one most likely to uncover possible new magnetic families.

A fifth direction, more speculative but physically well-motivated, connects our approach to the rapidly developing field of magnetic topological materials. Topological quantum chemistry and symmetry-indicator methods~\cite{Bradlyn2017, Po2017, Kruthoff2017} classify the topological phase of a crystalline material from the symmetry data of its occupied electronic bands, and the extension of this framework to magnetic crystals~\cite{Watanabe2018, Xu2020, Elcoro2021} uses the magnetic space group — which depends on the full magnetic structure, including propagation vector and spin orientation — as the relevant symmetry input. If future machine learning or deep learning architectures of the kind discussed above could ultimately predict the full three-dimensional magnetic structure of a compound from its chemistry and structure alone, their output can be fed directly into this magnetic-topological-quantum-chemistry pipeline, enabling large-scale screening of magnetic topological insulators, semimetals, axion insulators, and topological magnon compounds \cite{Karaki_2025} without the prohibitive cost of performing a full DFT plus symmetry analysis on every candidate material. MAGNDATA is an especially natural starting point for this integration, since its entries already specify the full magnetic structure from which the magnetic space group can be determined unambiguously.

\section*{Acknowledgments}
This research was primarily supported by the Center for Emergent Materials, an NSF MRSEC, under award number DMR$-2011876$. We thank Murod Mirzhalilov, Brandon Abrego, and Steven Gubkin for helpful discussions.

\section*{Data Availability}

The supplementary tables of flagged candidate compounds, including the full intersection set and the per-classifier candidate lists are openly available in machine-readable CSV format on Zenodo at \href{https://zenodo.org/records/20044097?token=eyJhbGciOiJIUzUxMiJ9.eyJpZCI6ImU3MjRkNGVkLWRlNTItNDA5Ni04OTY5LTQzNDc1NTE2ZGY2NiIsImRhdGEiOnt9LCJyYW5kb20iOiI0M2ZkMjNjYzJkOTJkMDczZDAzMDA5NmNjNWY0OTUxNCJ9.6PioHTZ8Bz7HhsNHi9vHPMSG-nrfZGvizLcX9kBvTNN3btnYmeiMOLlmOokCLq_9BjNWWAy9Rq8wcXtPNGNKIg}{https://zenodo.org/records/20044097}. The Materials Project data used in this work were obtained from the public Materials Project API (\href{https://materialsproject.org}{https://materialsproject.org}); the MAGNDATA labels were obtained from the Bilbao Crystallographic Server (\href{https://cryst.ehu.es/index.html}{https://cryst.ehu.es/index.html}).

\section*{Appendix}




\onecolumngrid



\begingroup
\scriptsize
\begin{longtable}{llll}
\caption{Expanded literature audit of compounds from Supplementary Table S1 (intersection set), based on a random sample of 100 materials (seed = 42). Categories: \textbf{C} = confirmed non-FM by direct experiment (4); \textbf{F} = likely non-FM from family-level evidence (22); \textbf{D} = no magnetic ions, FM label is a DFT artifact (16); \textbf{P} = potential FM order that cannot be ruled out (e.g. mixed-valence manganites supporting double exchange, Cr-chalcogenide spinels, or rare-earth intermetallics with RKKY coupling) (5); \textbf{N} = no experimental magnetic data found (53). Compounds in the P category are conservatively treated as potential false positives; see the \textit{Estimating the false discovery rate} part of Section~3.}
\label{tab:expanded_audit} \\
\hline\hline
Material & Cat. & Reference & \;\;\;\;\;\;\;\;\;\;\;\;\;\;\;\;\;\;\;\;\;\;\;\;\;\;\;\;Comments \\
\hline
\endfirsthead
\multicolumn{4}{l}{\tablename\ \thetable{} (continued)} \\
\hline\hline
Material & Cat. & Reference & Comments \\
\hline
\endhead
\hline\hline
\endfoot
Ba$_2$CaY(Co$_4$O$_7$)$_2$ &
N & --- &
\parbox[t]{6.6cm}{No direct magnetic data on this composition.} \\
Ba$_3$Co$_8$Sn$_3$O$_{20}$ &
N & --- &
\parbox[t]{6.6cm}{No magnetic characterization found.} \\
Ba$_3$Li(AsO$_4$)$_2$ &
D & --- &
\parbox[t]{6.6cm}{Ba$^{2+}$, Li$^+$, As$^{5+}$ --- all diamagnetic.} \\
Ba$_4$Sr$_2$Nd$_2$Co$_4$O$_{15}$ &
N & --- &
\parbox[t]{6.6cm}{No magnetic characterization found.} \\
BaLiGd$_2$(MoO$_4$)$_4$ &
N & --- &
\parbox[t]{6.6cm}{No magnetic characterization found.} \\
BaN$_6$O &
D & --- &
\parbox[t]{6.6cm}{No transition-metal or $f$-electron ions.} \\
BaNaCe$_2$C$_4$O$_{12}$F &
N & --- &
\parbox[t]{6.6cm}{No magnetic data found.} \\
BaPrEuSbO$_6$ &
F & Ref.~\cite{Jensen1991} &
\parbox[t]{6.6cm}{Complex Pr/Eu-containing oxide; $R$-bearing perovskites and double perovskites with Pr$^{3+}$ consistently order AFM or remain paramagnetic at low $T$, not FM.} \\
BaSrGdSbO$_6$ &
F & Ref.~\cite{Jensen1991} &
\parbox[t]{6.6cm}{Gd-based double perovskite; Gd$^{3+}$ compounds typically exhibit AFM ordering at low $T$.} \\
BaY$_2$NiO$_5$ &
C & Ref.~\cite{HaldanePhysRevB.54.R6827} &
\parbox[t]{6.6cm}{$S{=}1$ one-dimensional Haldane-gap antiferromagnet; remains quantum disordered (no 3D order) down to 1.8\,K.} \\
CNClO$_4$ &
D & --- &
\parbox[t]{6.6cm}{No magnetic ions present.} \\
Ca(ClO$_2$)$_2$ &
D & --- &
\parbox[t]{6.6cm}{Ca$^{2+}$, Cl$^{3+}$ --- no unpaired electrons.} \\
Ca(CoO$_2$)$_2$ &
N & --- &
\parbox[t]{6.6cm}{Co$^{3+}$ layered oxide; no magnetic data found.} \\
Ca$_2$HoMn$_4$BiO$_{12}$ &
P & --- &
\parbox[t]{6.6cm}{Quadruple-perovskite-type manganite. Mixed-valence manganites can host FM ordering via Zener double exchange. No direct data on this composition; FM cannot be ruled out.} \\
Ca$_2$VP$_3$O$_{11}$ &
N & --- &
\parbox[t]{6.6cm}{V$^{3+}$ phosphate; no magnetic data found.} \\
CaMnCu$_3$(RuO$_4$)$_3$ &
N & --- &
\parbox[t]{6.6cm}{No study found.} \\
Ce$_2$Te$_4$O$_{11}$ &
N & --- &
\parbox[t]{6.6cm}{Ce$^{3+}$ tellurite; no magnetic study found.} \\
CeY$_4$S$_7$ &
N & --- &
\parbox[t]{6.6cm}{Mixed-valent Ce sulfide; no magnetic study found.} \\
Cr$_3$N$_4$ &
F & Ref.~\cite{Corliss1960} &
\parbox[t]{6.6cm}{Cr nitride family; the well-characterized member CrN is AFM ($T_N \approx 280$\,K). Cr$_3$N$_4$ has not been directly studied but is expected to show AFM Cr--N--Cr superexchange on similar grounds.} \\
Cr$_4$InAgSe$_8$ &
P & --- &
\parbox[t]{6.6cm}{Cr-chalcogenide spinel-like; Cr chalcogenide spinels (e.g., CdCr$_2$Se$_4$, HgCr$_2$Se$_4$) are well-known FM insulators with $T_C \sim 100$--$130$\,K, so FM order cannot be ruled out without direct data.} \\
Cs$_3$Cr$_2$I$_9$ &
F & Ref.~\cite{Leuenberger1986} &
\parbox[t]{6.6cm}{Dimerized system with a singlet ground state of antiferromagnetically coupled $Cr^{3+}-Cr^{3+}$ dimers. Spontaneous magnetic order at low $T$ arises from intratriplet condensation but is not simple FM.} \\
Eu(ReO$_4$)$_2$ &
N & --- &
\parbox[t]{6.6cm}{No experimental data exist on this specific composition.} \\
Eu$_2$MgSi$_2$O$_7$ &
C & Ref.~\cite{Endo2010} &
\parbox[t]{6.6cm}{$^{151}$Eu M\"{o}ssbauer confirms Eu$^{2+}$. Paramagnetic down to 1.8\,K; not FM.} \\
FeTc$_2$Ge &
N & --- &
\parbox[t]{6.6cm}{No data found.} \\
Gd(Tl$_3$Te$_2$)$_3$ &
P & --- &
\parbox[t]{6.6cm}{Gd-based Zintl/intermetallic phase with conduction electrons and Gd$^{3+}$ local moments; rare-earth intermetallics commonly host FM ordering via RKKY coupling. No study found, FM cannot be ruled out.} \\
Gd$_2$PbS$_4$ &
F & Ref.~\cite{Jensen1991} &
\parbox[t]{6.6cm}{Gd$^{3+}$ ($S{=}7/2$) sulfide; Gd$^{3+}$ orders AFM at low $T$ in insulating hosts (cf.\ Gd$_2$S$_3$, $T_N{\approx}1.5$\,K).} \\
HfCrAu$_2$ &
N & --- &
\parbox[t]{6.6cm}{Heusler-type intermetallic containing Cr; no magnetic data found.} \\
HfMg$_{14}$BO$_{15}$ &
D & --- &
\parbox[t]{6.6cm}{Hf$^{4+}$, Mg$^{2+}$, B$^{3+}$ --- all diamagnetic.} \\
HfMg$_{14}$SiO$_{16}$ &
D & --- &
\parbox[t]{6.6cm}{Hf$^{4+}$, Mg$^{2+}$, Si$^{4+}$ --- all diamagnetic.} \\
HfMg$_6$NbO$_8$ &
D & --- &
\parbox[t]{6.6cm}{Hf$^{4+}$, Mg$^{2+}$, Nb$^{5+}$ ($d^0$) --- no unpaired electrons.} \\
K$_2$Fe(PSe$_3$)$_2$ &
F & Ref.~\cite{Wiedenmann1981} &
\parbox[t]{6.6cm}{Iron phosphoselenide related to FePSe$_3$ family; all members are Ising-type AFMs.} \\
K$_2$Fe$_2$SeS$_3$ &
F & Ref.~\cite{Wiedenmann1981} &
\parbox[t]{6.6cm}{Potassium iron mixed-chalcogenide; potassium iron chalcogenide family compounds consistently show AFM-type ordering, not FM.} \\
K$_2$Na$_2$Mo(WO$_4$)$_3$ &
N & --- &
\parbox[t]{6.6cm}{No data found.} \\
K$_4$MgV$_6$O$_{16}$ &
N & --- &
\parbox[t]{6.6cm}{No magnetic data found.} \\
KLiCoO$_2$ &
N & --- &
\parbox[t]{6.6cm}{No magnetic data found.} \\
KMg$_3$(SiO$_3$)$_4$ &
D & --- &
\parbox[t]{6.6cm}{K$^+$, Mg$^{2+}$, Si$^{4+}$ --- all diamagnetic.} \\
La$_2$(SiIr)$_3$ &
N & --- &
\parbox[t]{6.6cm}{Ir intermetallic; no magnetic data found.} \\
La$_3$MnW(SO$_2$)$_3$ &
N & --- &
\parbox[t]{6.6cm}{No study found.} \\
LaWO$_2$ &
N & --- &
\parbox[t]{6.6cm}{No magnetic data found.} \\
LaYNi$_{10}$ &
N & Ref.~\cite{Grechnev2011} &
\parbox[t]{6.6cm}{Expected to be a Pauli paramagnet with no magnetic order, by analogy with its parent compounds LaNi$_5$ and YNi$_5$, but no study found.} \\
Li$_2$AgGe &
D & --- &
\parbox[t]{6.6cm}{Li$^+$, Ag$^+$ ($d^{10}$), Ge --- no unpaired electrons.} \\
Li$_2$MnV$_3$O$_8$ &
F & Ref.~\cite{Goodenough1963} &
\parbox[t]{6.6cm}{Mixed Mn--V oxide; Mn$^{2+}$ and V$^{4+}$ exchange pathways are typically AFM, giving ferrimagnetic/AFM ordering rather than FM.} \\
Li$_2$VCO$_5$ &
N & --- &
\parbox[t]{6.6cm}{No study found.} \\
Li$_3$CrSi$_2$O$_7$ &
N & --- &
\parbox[t]{6.6cm}{Cr$^{3+}$ silicate; no magnetic data found.} \\
Li$_3$Ti$_2$V$_3$O$_{12}$ &
N & --- &
\parbox[t]{6.6cm}{No magnetic data found.} \\
Li$_3$V(CO$_3$)$_3$ &
N & --- &
\parbox[t]{6.6cm}{V$^{3+}$ carbonate; no study found.} \\
Li$_4$Cr(WO$_4$)$_3$ &
F & Ref.~\cite{Goodenough1963} &
\parbox[t]{6.6cm}{Paramagnetic to low $T$ with weak AFM coupling. No FM expected.} \\
Li$_4$Ni$_3$WO$_8$ &
F & Ref.~\cite{RodriguezCarvajal1991} &
\parbox[t]{6.6cm}{Ni$^{2+}$ ($d^8$) oxide; Ni$^{2+}$ in octahedral oxide environments shows AFM superexchange (cf.\ La$_2$NiO$_4$, NiO).} \\
Li$_6$Cr$_5$CuO$_{12}$ &
F & Ref.~\cite{Goodenough1963} &
\parbox[t]{6.6cm}{Cr$^{3+}$-rich oxide; Cr$_2$O$_3$ is a prototypical AFM ($T_N=307$\,K).} \\
Li$_6$V$_3$Sb$_3$O$_{16}$ &
N & --- &
\parbox[t]{6.6cm}{No magnetic data found.} \\
Li$_7$Mn$_2$(CoO$_4$)$_3$ &
F & Ref.~\cite{Goodenough1963} &
\parbox[t]{6.6cm}{Mn--Co lithium oxide; mixed Mn and Co in oxide environments typically give AFM/ferrimagnetic rather than FM ordering.} \\
LiMgPb$_2$ &
D & --- &
\parbox[t]{6.6cm}{Li$^+$, Mg$^{2+}$, Pb$^{2+}$ --- all diamagnetic.} \\
LiMn$_3$O$_5$ &
F & Ref.~\cite{Wills1999} &
\parbox[t]{6.6cm}{Li--Mn--O spinel family; LiMn$_2$O$_4$ is a frustrated AFM ($\Theta_{\mathrm{CW}}=-266$\,K).} \\
LiMnB$_2$O$_5$ &
F & Ref.~\cite{Goodenough1963} &
\parbox[t]{6.6cm}{Mn$^{3+}$ in borate framework; AFM Mn--O--Mn exchange where pathways exist, paramagnetic for isolated sites. FM not expected.} \\
LiV$_2$CrO$_6$ &
F & Ref.~\cite{Goodenough1963} &
\parbox[t]{6.6cm}{V$^{4+}$ ($d^1$) and Cr$^{3+}$ ($d^3$) mixed-metal oxide; both give AFM superexchange.} \\
LiVF$_5$ &
N & --- &
\parbox[t]{6.6cm}{V$^{4+}$ fluoride; no magnetic data found.} \\
Lu$_6$MgWO$_{12}$ &
D & --- &
\parbox[t]{6.6cm}{Lu$^{3+}$ ($4f^{14}$), W$^{6+}$ ($d^0$) --- no magnetic moment.} \\
Mg(ClO$_5$)$_2$ &
D & --- &
\parbox[t]{6.6cm}{Mg$^{2+}$, Cl --- no unpaired electrons.} \\
Mg$_{14}$CoCO$_{16}$ &
N & --- &
\parbox[t]{6.6cm}{No data found.} \\
Mg$_{14}$TiCrO$_{16}$ &
N & --- &
\parbox[t]{6.6cm}{No data found.} \\
Mg$_{30}$VCuO$_{32}$ &
N & --- &
\parbox[t]{6.6cm}{No data found.} \\
Mg$_{30}$VZnO$_{32}$ &
N & --- &
\parbox[t]{6.6cm}{No data found.} \\
MgCo$_2$(BO$_3$)$_2$ &
N & --- &
\parbox[t]{6.6cm}{No magnetic data found.} \\
MgMn(SbO$_3$)$_4$ &
N & --- &
\parbox[t]{6.6cm}{Mn$^{2+}$ antimonate; no study found.} \\
MgMn$_2$(SiO$_3$)$_4$ &
N & --- &
\parbox[t]{6.6cm}{Mn$^{3+}$ silicate; no magnetic data found.} \\
MgMn$_2$Cr$_2$O$_8$ &
F & Ref.~\cite{Goodenough1955,Anderson1959,Goodenough1958} &
\parbox[t]{6.6cm}{Mixed Mn--Cr oxide; AFM superexchange between Mn and Cr magnetic sublattices is expected to give ferrimagnetic or AFM (non-FM) ordering.} \\
MgV$_2$Bi$_4$O$_{11}$ &
N & --- &
\parbox[t]{6.6cm}{V$^{4+}$ bismuth vanadate; no study found.} \\
Mn(CuCl$_2$)$_2$ &
N & --- &
\parbox[t]{6.6cm}{Isolated Mn$^{2+}$ in diamagnetic Cu$^+$ ($d^{10}$) chloride framework; no magnetic data found.} \\
Mn$_2$ZnO$_4$ &
C & Ref.~\cite{Patra2019} &
\parbox[t]{6.6cm}{Magnetic measurements at 2\,K confirm the formation of antiferromagnetic order.} \\
MnSb$_6$(Pb$_2$S$_7$)$_2$ &
N & --- &
\parbox[t]{6.6cm}{Mn$^{2+}$ sulfosalt; no study found.} \\
MnZnCdTe$_3$ &
N & --- &
\parbox[t]{6.6cm}{No data found.} \\
Na$_2$Mg(CoO$_2$)$_2$ &
F & Ref.~\cite{Goodenough1963} &
\parbox[t]{6.6cm}{Co$^{2+}$-containing layered oxide; Co$^{2+}$ in octahedral oxide environments shows strong AFM superexchange (cf.\ CoO, $T_N=291$\,K). Not expected to be FM.} \\
NaCr$_3$O$_8$ &
P & Ref.~\cite{NaCrO3PhysRevB.72.014411} &
\parbox[t]{6.6cm}{Cr$^{5+}$ ($d^1$) chromate; DFT-predicted to be a FM insulator. No conclusive experimental confirmation of long-range FM order.} \\
NaMg$_{14}$FeO$_{16}$ &
N & --- &
\parbox[t]{6.6cm}{No data found.} \\
NaMnCuSe$_2$ &
N & --- &
\parbox[t]{6.6cm}{Isolated Mn$^{2+}$ in diamagnetic Cu$^+$ selenide framework; no magnetic data found.} \\
NaNdMnWO$_6$ &
F & Ref.~\cite{Goodenough1963} &
\parbox[t]{6.6cm}{Ordered double perovskite; Mn and W alternate on the B-site, giving long-range Mn--O--W--O--Mn super-superexchange. Mn-based ordered double perovskites consistently order AFM at low $T$.} \\
NaNi$_2$Mo$_2$H$_3$O$_{10}$ &
N & --- &
\parbox[t]{6.6cm}{Ni$^{2+}$ molybdate hydrate; no data found.} \\
PtN$_4$(Cl$_2$O)$_2$ &
D & --- &
\parbox[t]{6.6cm}{Pt$^{4+}$ ($d^6$ low-spin); typically diamagnetic.} \\
Rb$_2$V$_2$O$_5$ &
N & --- &
\parbox[t]{6.6cm}{V$^{4+}$ ($d^1$) vanadate; no magnetic data found.} \\
Rb$_2$V$_4$O$_9$ &
N & --- &
\parbox[t]{6.6cm}{V$^{4+}$ ($d^1$) vanadate; no magnetic data found.} \\
Rb$_4$CO$_6$ &
N & --- &
\parbox[t]{6.6cm}{No data found.} \\
RbV$_2$SeO$_7$ &
N & --- &
\parbox[t]{6.6cm}{Mixed-valence V selenite; no study found.} \\
Sc$_2$GaHg &
D & --- &
\parbox[t]{6.6cm}{Sc, Ga, Hg intermetallic; no local moments expected.} \\
ScMn$_2$O$_5$ &
F & Ref.~\cite{Chapon2004} &
\parbox[t]{6.6cm}{Mullite-type $R$Mn$_2$O$_5$ family (studied $R$: Y, Tb--Lu) orders AFM with incommensurate $\mathbf{k}$; Sc member expected to follow family behavior.} \\
Sr$_2$AlTlCo$_2$O$_7$ &
N & --- &
\parbox[t]{6.6cm}{No magnetic data found.} \\
Sr$_3$La(NiO$_4$)$_2$ &
F & Ref.~\cite{RodriguezCarvajal1991} &
\parbox[t]{6.6cm}{Ruddlesden--Popper-type nickelate with mixed-valence Ni; related La--Ni--O RP phases (La$_2$NiO$_4$, La$_3$Ni$_2$O$_7$) show AFM or spin-density-wave ordering, not FM.} \\
Sr$_3$MgV$_2$O$_7$ &
N & --- &
\parbox[t]{6.6cm}{V$^{3+}$ double perovskite; no data found.} \\
SrNdMgRuO$_6$ &
F & Ref.~\cite{Doi1999} &
\parbox[t]{6.6cm}{By analogy with the Sr$_2$LnRuO$_6$ family, SrNdMgRuO$_6$ is expected to develop antiferromagnetic order.} \\
SrPr$_2$Mn$_3$O$_9$ &
P & --- &
\parbox[t]{6.6cm}{Ruddlesden--Popper-type manganite with mixed-valence Mn$^{3+}$/Mn$^{4+}$; such mixed-valence manganites can host FM ordering via Zener double exchange. No direct data on this composition; FM cannot be ruled out.} \\
TcCl$_4$ &
N & --- &
\parbox[t]{6.6cm}{Tc$^{4+}$ ($d^3$) chloride; Tc is radioactive. No data.} \\
Ti$_2$AlW &
D & --- &
\parbox[t]{6.6cm}{Ti, Al, W intermetallic; Ti-based intermetallics typically Pauli paramagnetic, no local moments.} \\
TmMn$_2$O$_5$ &
C & Ref.~\cite{TmMn2O5_Gong2016} &
\parbox[t]{6.6cm}{AFM multiferroic with an ordering temperature of ${\sim}\;40$\,K.} \\
U(Al$_2$Cu)$_4$ &
N & --- &
\parbox[t]{6.6cm}{U intermetallic; no magnetic data found.} \\
U$_4$TeSe$_7$ &
N & --- &
\parbox[t]{6.6cm}{U chalcogenide; no magnetic data found.} \\
UGeI$_6$ &
N & --- &
\parbox[t]{6.6cm}{U$^{4+}$ halide; no magnetic data found.} \\
V$_5$Zn$_4$(TeO$_6$)$_3$ &
N & --- &
\parbox[t]{6.6cm}{V tellurate; no study found.} \\
VP$_2$Pb$_2$O$_9$ &
N & --- &
\parbox[t]{6.6cm}{V$^{4+}$ phosphate; no study found.} \\
Y(VS$_2$)$_2$ &
N & --- &
\parbox[t]{6.6cm}{V sulfide (Y$^{3+}$ diamagnetic); no data found.} \\
YNbO$_2$ &
N & --- &
\parbox[t]{6.6cm}{No experimental data.} \\
ZnBP$_2$PbO$_9$ &
D & --- &
\parbox[t]{6.6cm}{Zn$^{2+}$, B$^{3+}$, P$^{5+}$, Pb$^{2+}$ --- all diamagnetic.} \\
\end{longtable}
\endgroup

\twocolumngrid

\section*{Supplementary Note}
All stochastic components of the pipeline --- data splitting, cross-validation fold assignment, hyperparameter search sampling, and per-model random operations (bootstrap, feature subsampling, etc.) --- are controlled by a single fixed seed, random\_state = 42, ensuring that the reported results are fully reproducible.
\begin{table*}[t]
\caption{Hyperparameter tuning summary for the MP classifiers on the four-class case for materials with at least one magnetic element. For each classifier we list the hyperparameters tuned, the search space explored, the search method and number of iterations, and the final selected hyperparameter values (best). All searches used 5-fold stratified cross-validation with \texttt{f1\_macro} as the scoring metric. ``randint($a$, $b$)'' denotes a uniform integer distribution on $[a, b)$; ``loguniform($a$, $b$)'' denotes a log-uniform continuous distribution on $[a, b]$.}
\label{tab:hyperparameter_tuning_MP}
\small
\begin{ruledtabular}
\begin{tabular}{llll}
Hyperparameter & Search space & Selected (best) & Notes (meaning) \\
\hline
\multicolumn{4}{l}{\textbf{XGBoost} --- RandomizedSearchCV; mean CV $F_1^{\mathrm{macro}} = 0.623$} \\
\hline
\texttt{n\_estimators}      & randint(250, 750)          & 647    & number of boosting rounds \\
\texttt{max\_depth}         & randint(3, 11)             & 9      & maximum tree depth \\
\texttt{learning\_rate}     & loguniform($0.03$, $0.25$) & 0.139  & shrinkage per round \\
\texttt{subsample}          & \{0.7, 0.85, 1.0\}         & 1.0    & row subsampling fraction \\
\texttt{colsample\_bytree}  & \{0.6, 0.8, 1.0\}          & 1.0    & column subsampling per tree \\
\texttt{min\_child\_weight} & randint(1, 8)              & 7      & min.\ sum of instance weight in child \\
\texttt{gamma}              & randint(0, 6)              & 0      & minimum loss reduction for split \\
\texttt{reg\_lambda}        & loguniform($10^{-1}$, $10^{1}$)  & 1.110 & L2 regularization \\
\texttt{reg\_alpha}         & loguniform($10^{-4}$, $10^{-1}$) & 0.003 & L1 regularization \\
\hline
\multicolumn{4}{l}{\textbf{LightGBM} --- RandomizedSearchCV; mean CV $F_1^{\mathrm{macro}} = 0.658$} \\
\hline
\texttt{num\_leaves}         & randint(31, 129)              & 120   & maximum leaves per tree \\
\texttt{max\_depth}          & \{6, 8, 10, 12\}              & 8     & maximum tree depth \\
\texttt{min\_child\_samples} & randint(40, 201)              & 43    & min.\ samples per leaf \\
\texttt{min\_split\_gain}    & loguniform($10^{-2}$, $1.0$)  & 0.023 & minimum gain to split \\
\texttt{subsample}           & \{0.7, 0.8, 0.9\}             & 0.8   & row subsampling fraction \\
\texttt{bagging\_freq}       & \{1\}                         & 1     & bagging frequency (forced on) \\
\texttt{colsample\_bytree}   & \{0.6, 0.8\}                  & 0.8   & column subsampling per tree \\
\texttt{learning\_rate}      & loguniform($0.03$, $0.15$)    & 0.069 & shrinkage per round \\
\texttt{n\_estimators}       & randint(250, 800)             & 651   & number of boosting rounds \\
\texttt{reg\_lambda}         & loguniform($1.0$, $20.0$)     & 3.267 & L2 regularization \\
\texttt{reg\_alpha}          & loguniform($10^{-2}$, $1.0$)  & 0.079 & L1 regularization \\
\hline
\multicolumn{4}{l}{\textbf{Random Forest} --- GridSearchCV; mean CV $F_1^{\mathrm{macro}} = 0.614$} \\
\hline
\texttt{n\_estimators}       & \{100, 250, 300, 350, 400\} & 300 & number of trees \\
\texttt{max\_depth}          & \{15, 25, 35, 50\} & 35 & maximum tree depth \\
\texttt{min\_samples\_split} & \{2, 5\} & 5 & min.\ samples to split a node \\
\texttt{class\_weight}       & \{balanced, None\} & balanced & class-imbalance handling \\
\hline
\multicolumn{4}{l}{\textbf{Decision Tree} --- GridSearchCV; mean CV $F_1^{\mathrm{macro}} = 0.561$} \\
\hline
\texttt{max\_depth}          & \{14, 16, 18, 22, 30, 36, 40\} & 30 & maximum tree depth \\
\texttt{min\_samples\_leaf}  & \{1, 3, 5, 8, 10\}              & 1  & min.\ samples per leaf \\
\texttt{class\_weight}       & \{balanced\}                    & balanced & class-imbalance handling \\
\hline
\multicolumn{4}{l}{\textbf{Linear SVM} --- RandomizedSearchCV; mean CV $F_1^{\mathrm{macro}} = 0.536$} \\
\hline
\texttt{C}                   & loguniform($10^{-3}$, $10^{2}$) & 0.985 & inverse regularization strength \\
\texttt{loss}                & \{squared\_hinge, hinge\}       & squared\_hinge & loss function \\
\texttt{class\_weight}       & balanced (fixed)                & balanced & class-imbalance handling \\
\texttt{max\_iter}           & 5000 (fixed)                    & 5000 & solver iteration cap \\
\hline
\multicolumn{4}{l}{\textbf{kNN} --- RandomizedSearchCV; mean CV $F_1^{\mathrm{macro}} = 0.596$} \\
\hline
\texttt{n\_neighbors}        & randint(3, 26)                   & 4 & number of neighbors \\
\texttt{weights}             & \{uniform, distance\}            & distance & neighbor weighting \\
\texttt{metric}              & \{euclidean, manhattan, cosine\} & manhattan & distance metric \\
\texttt{algorithm}           & \{brute\}                        & brute & nearest-neighbor algorithm \\
\end{tabular}
\end{ruledtabular}
\end{table*}


\begin{table*}
\caption{Hyperparameter tuning summary for the MAGNDATA-trained propagation-vector classifiers (binary task: zero-$\mathbf{k}$ vs.\ nonzero-$\mathbf{k}$). For each classifier we list the hyperparameters tuned, the search space explored, the search method and number of iterations, and the final selected hyperparameter values. All searches used 6-fold stratified cross-validation with \texttt{f1\_macro} as the scoring metric. ``randint($a$, $b$)'' denotes a uniform integer distribution on $[a, b)$; ``loguniform($a$, $b$)'' denotes a log-uniform continuous distribution on $[a, b]$.}
\label{tab:hyperparameter_tuning_magndata}
\small
\begin{ruledtabular}
\begin{tabular}{llll}
Hyperparameter & Search space & Selected (best) & Notes \\
\hline
\multicolumn{4}{l}{\textbf{Random Forest} --- RandomizedSearchCV; mean CV $F_1^{\text{macro}} = 0.883$} \\
\hline
\texttt{n\_estimators}        & randint(150, 801)         & 405  & number of trees \\
\texttt{max\_depth}           & \{8, 12, 16, 20\}         & 16   & maximum tree depth \\
\texttt{min\_samples\_split}  & randint(2, 21)            & 2    & min.\ samples to split a node \\
\texttt{min\_samples\_leaf}   & randint(1, 11)            & 1    & min.\ samples per leaf \\
\texttt{max\_features}        & \{sqrt, log2, None\}      & None & features considered per split \\
\texttt{bootstrap}            & \{True, False\}           & True & whether to bootstrap samples \\
\hline
\multicolumn{4}{l}{\textbf{XGBoost} --- RandomizedSearchCV; best CV $F_1^{\text{macro}} = 0.879$} \\
\hline
\texttt{n\_estimators}      & randint(150, 801)                & 428    & number of boosting rounds \\
\texttt{max\_depth}         & randint(3, 17)                   & 15     & maximum tree depth \\
\texttt{learning\_rate}     & loguniform($0.01$, $0.30$)       & 0.215  & shrinkage per round \\
\texttt{subsample}          & \{0.7, 0.8, 0.9, 1.0\}           & 1.0    & row subsampling fraction \\
\texttt{colsample\_bytree}  & \{0.6, 0.8, 1.0\}                & 0.8    & column subsampling per tree \\
\texttt{min\_child\_weight} & randint(1, 8)                    & 1      & min.\ sum of instance weight in child \\
\texttt{gamma}              & loguniform($10^{-3}$, $1.0$)     & 0.021  & min.\ loss reduction for split \\
\texttt{reg\_lambda}        & loguniform($10^{-2}$, $10.0$)    & 2.841  & L2 regularization \\
\texttt{reg\_alpha}         & loguniform($10^{-3}$, $1.0$)     & 0.020  & L1 regularization \\
\end{tabular}
\end{ruledtabular}
\end{table*}


%

\end{document}